\documentclass{IEEEtran}
\usepackage{graphicx}
\usepackage{placeins}
\usepackage{float}
\usepackage{hyperref}
\usepackage{tabularx,colortbl}
\usepackage{soul}
\usepackage{xcolor}
\usepackage{balance}
\usepackage{amssymb}
\usepackage{amsmath}
\usepackage{amsfonts}
\usepackage{epsfig}
\usepackage{psfrag}
\usepackage{epstopdf}
\usepackage[mode=errorstop]{pstool}
\usepackage{subfigure}
\usepackage{cite}

\begin{document}
\title{General Mapping between \\ Complex Spatial and Temporal Frequencies \\ by Analytical Continuation}

\author{Mojtaba Dehmollaian, \IEEEmembership{Senior Member, IEEE} and Christophe Caloz, \IEEEmembership{Fellow, IEEE}
	\thanks{The paper is submitted on 2 June 2020. \newline
		M. Dehmollaian is with the Poly-Grames Research Center, Polytechnique Montr\'{e}al, Montr\'{e}al, Qu\'{e}bec (e-mail: mojtaba.dehmollaian@polymtl.ca). \newline C. Caloz is with the ESAT-TELEMIC Research Center, KU Leuven, Leuven, Belgium (e-mail: christophe.caloz@kuleuven.be).}}

\maketitle

\begin{abstract}
This paper introduces a general technique for inter-mapping the complex spatial frequency (or propagation constant) $\gamma=\alpha+j\beta$ and the temporal frequency $\omega = \omega_\text{r}+j\omega_\text{i}$ of an arbitrary electromagnetic structure. This technique, based on the analytic property of complex functions describing physical phenomena, invokes the analytic continuity theorem to assert the unicity of the mapping function, and find this function from known data within a restricted domain of its analycity by curve-fitting to a generic polynomial expansion. It is not only applicable to canonical problems admitting an analytical solution, but to \emph{any} problems, from eigen-mode or driven-mode full-wave simulation results. The proposed technique is demonstrated for several systems, namely an unbounded lossy medium, a dielectric-filled rectangular waveguide, a periodically-loaded transmission line, a one-dimensional photonic crystal and a series-fed patch (SFP) leaky-wave antenna (LWA), and it is validated either by analytical results or by full-wave simulated results.
\end{abstract}
\IEEEpeerreviewmaketitle
\section{Introduction}\label{sec:intro}
\IEEEPARstart{W}{aves} propagate in both space and time. They are therefore characterized by both spatial and temporal frequencies~\cite{maxwell1865viii,jackson2007classical}. The spatial frequency, $\mathbf{k}$ (vector), or inverse space, and the temporal frequency, $\omega$ (scalar), or inverse time, are the Fourier counterparts of the position, $\mathbf{r}$ (vector), or direct space, and time, $t$ (scalar), or direct time, respectively, where the terms ‘direct’ and ‘inverse’ refer to the independent variables of the usual Fourier transform-pair~\cite{ishimaru2017electromagnetic,joseph2017introduction}. The spatial and temporal spectra are properties of waves that are as fundamental as their direct-space features~\cite{joseph2017introduction,pozar2011microwave,Caloz_TAP1_02_2020,Caloz_TAP2_02_2020}.

Although these spatial and temporal frequencies are generally complex, and may therefore be written as
\begin{subequations}\label{eq:all_cpl}
	\begin{equation}\label{eq:all_k}
	k=\beta-j\alpha=-j\gamma
	\end{equation}
	and
	\begin{equation}\label{eq:all_w}
	\omega=\omega_\text{r}+j\omega_\text{i}=2\pi(f_\text{r}+jf_\text{i}),
	\end{equation}	
\end{subequations}
where we have temporarily reduced the vector $\mathbf{k}$ to a scalar, $k$, for simplicity, they are most often \emph{not simultaneously} complex\footnote{There are exceptions, such as lossy photonic crystals~\cite{Joannopoulos1995photonic} and all spacetime crystals~\cite{Deck_APH_10_2019}, where the two frequencies are simultaneously complex.}. Depending on the excitation, we typically have $\omega$ real and $k$ complex or $k$ real and $\omega$ complex. In the Traveling-Wave (TW) regime, as for instance for propagation in an unbounded medium or in a matched waveguide, the temporal frequency is purely real and the spatial frequency is complex, i.e., 
\begin{subequations}\label{eq:w_re}
	\begin{equation}
	\omega=\omega_\text{r}
	\end{equation}
	and
	\begin{equation}
	k=\beta-j\alpha,
	\end{equation}	
\end{subequations}
while in the Standing-Wave (SW) regime, as for instance for propagation in a bounded medium, in a terminated waveguide or in a resonant scatterer, the spatial frequency is purely real and the temporal frequency is complex, i.e.,
\begin{subequations}\label{eq:k_re}
	\begin{equation}
	k=\beta
	\end{equation}
	and
	\begin{equation}
	\omega=\omega_\text{r}+j\omega_\text{i}.
	\end{equation}	
\end{subequations}

The difference between the natures of the frequencies involved in the TW regime [Eq.~\eqref{eq:w_re}] and in SW regime [Eq.~\eqref{eq:k_re}], raises the following \emph{fundamental question}: Can one, for a given system, systematically find the \emph{correspondence} between the TW-regime quantities ($\omega_\text{r}$, $\beta$, $\alpha$) and the SW-regime quantities ($\beta$, $\omega_\text{r}$, $\omega_\text{i}$), and vice versa? The answer to this question is positive, and it is the object of this paper to demonstrate this fact and to provide a \emph{mapping technique} applicable to \emph{any} electromagnetic system.

In fact, this question originally occurred to the mind of the second author, about 15 years ago, in connection with periodic leaky-wave antenna (LWA) structures~\cite{jackson2012leaky}. Whereas a periodic LWA is typically operated in the TW-regime, and hence characterized by the triplet ($\omega_\text{r}$, $\beta$, $\alpha$), it is most conveniently analyzed in the SW-regime, or more precisely in the Periodic Boundary Condition (PBC) regime, characterized by the triplet ($\beta$, $\omega_\text{r}$, $\omega_\text{i}$), with $\omega_\text{i}\neq{0}$ in the fast-wave region of the dispersion diagram, even in the absence of dissipative loss, due to leakage. How could one obtain the useful former triplet, where $\alpha$ represents the leakage factor of the antenna, from the computed latter triplet, where $\omega_\text{i}=1/\tau=\omega_0/(2Q)$, with $\tau$ being the relaxation time, $\omega_0$ the resonance frequency and $Q$ the related unloaded quality factor~\cite{desoer2010basic}?

In the periodic LWA problem, the dispersion parameters, ($\omega_\text{r}$, $\beta$, $\alpha$), of the infinite structure or, equivalently, of the matched structure, may be computed in the TW-regime by simulating the actual periodic LWA structure as a two-port network with $N$ unit cells, without any dissipative loss, so that all the simulated loss accounts for radiation only, and computing $\beta$ and $\alpha$ versus the excitation frequency, $\omega_\text{r}$, as $\beta=-\varphi^\text{uw}\left\{S_{21}\right\}/\ell$ and $\alpha=-\ln(|S_{21}|/\sqrt{1-|S_{11}|^2})/\ell$, respectively, where $S_{21}$ and $S_{11}$ are the transmission and reflection scattering parameters of the two-port  waveguiding structure, $\varphi^\text{uw}$ denotes the unwrapped phase~\cite{Caloz_Wiley_2006}, and $\ell=Np$, with $p$ being the period, is the total length of the structure\footnote{A LWA structure under design generally has an unknown and dispersive impedance. Therefore, it has a frequency-dependent mismatch ($|S_{11}(\omega)|\neq{0}$) in the typically constant-impedance (e.g. $50~\Omega$) of the simulation environment. The corresponding return loss, $|S_{11}(\omega)|^2$, is unrelated to the radiation of the antenna, and must therefore be extracted from the simulated scattering parameters for a proper estimation of its propagation constant and leakage factor. This is accomplished by accounting for the fact that the power penetrating into the LWA structure is $1-|S_{11}(\omega)|^2$ smaller than the power at the input port, i.e., $S_{21}^\text{str}=S_{21}\sqrt{1-|S_{11}(\omega)|^2}$, noting that $S_{21}^\text{str}=\text{e}^{-j(\beta-j\alpha)\ell}$, equating the last two results, and respectively solving for $\beta$ and for $\alpha$, which leads to the formulas given in the text.}. Such a \emph{driven-mode} procedure, used for instance in~\cite{Otto_TAP_10_2014}, is very inefficient, not only because it requires the simulation of $N$ identical cells, but also because it requires increasing the number of cells as $N\rightarrow{N+1}$ until the computed $\beta(\omega)$ and $\alpha(\omega)$ have converged to their final periodic value\footnote{A LWA structure is typically several wavelengths long and has therefore essentially the same spatial spectrum as its infinite counterpart~\cite{jackson2012leaky}.} In contrast, the PBC-regime \emph{eigen-mode} problem only necessitate the computation of the (unique) periodic cell, which is much faster, but, unfortunately, delivers the awkward triplet ($\beta$, $\omega_\text{r}$, $\omega_\text{i}$) instead of the desired triplet ($\omega_\text{r}$, $\beta$, $\alpha$)!

A relation between the complex temporal and spatial frequencies for periodic structures was first reported in~\cite{otto2012complex}, and an attempt to find a general mapping solution, starting with the simplest problem of plane-wave propagation in an unbounded lossy medium, was discussed in~\cite{dyab2015interpretation}. The topic was recently revisited in~\cite{king2019relation}, where a mapping relation was derived for the problem of a (closed) waveguide. However, all these studies are restricted to problems having a \emph{known analytical solution}. Such solutions are therefore simple, because they just result from inverting an analytical function, but they are of limited practical interest since only a few canonical structures admit an analytical solution, whereas virtually all practical engineering structures can be simulated only numerically.

This paper resolves the long-lasting problem of the \emph{general mapping} between the complex spatial and temporal frequencies of electromagnetic structures, without requiring an analytical solution. It shows that, based on the theorem of analytical continuation and on the physical nature of real phenomena, the mapping function between the two complex spaces is necessarily \emph{unique}, and presents a general technique for determining and inverting this function, using a general polynomial approximation of the function with fitted parameters. This method applies to any structure, periodic or nonperiodic, with simulated triplets ($\omega_\text{r}$, $\beta$, $\alpha$) or ($\beta$, $\omega_\text{r}$, $\omega_\text{i}$), and systematically allows to convert one to the other.

The paper is organized as follows. Section~\ref{sec:prob_stat} explains the concept of complex frequencies (usual frequency and propagation constant) and the problem. Section~\ref{sec:spec_examples} gives examples of problems with analytical solutions, as later benchmarks for the proposed method. Section~\ref{sec:General_solution} presents the proposed general mapping technique, based on the analytical continuation theorem and polynomial curve-fitting. Section~\ref{sec:val_ill} provides a numerical validation and illustration of the method for various examples, including the periodic LWA at the origin of this research. Finally, Sec.~\ref{sec:Dis_Con} closes the paper.

\section{Statement of the Problem}\label{sec:prob_stat}

This section first describes the complex spatial frequency [Eq.~\eqref{eq:all_k}] and the complex temporal frequency [Eq.~\eqref{eq:all_w}], and then states the general plane mapping problem between them.

\subsection{Complex Spatial and Temporal Frequencies}\label{sec:concept_cf_cg}
\begin{figure}[h!]
	\includegraphics[]{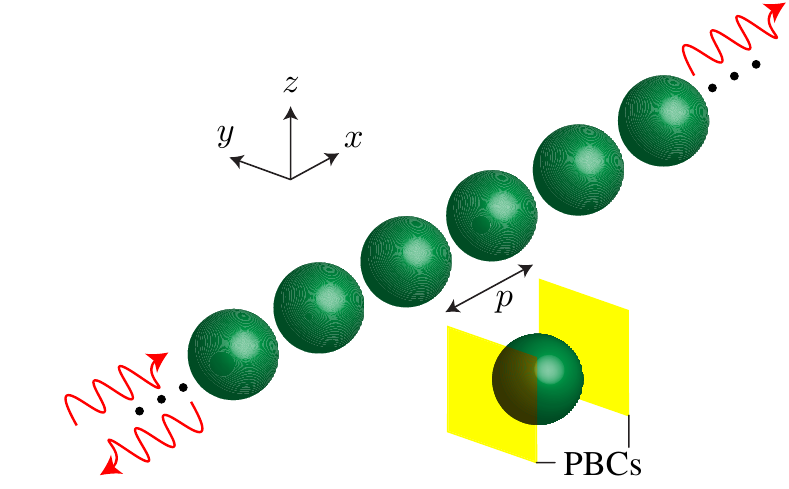}
	\caption{Generic one-dimensional periodic structure.}\label{fig:gperiodic}
\end{figure}

We shall consider here a periodic problem, because such a problem is also a generalization of a non-periodic problem and because it best relates to the original problem of the periodic LWA. We assume a one-dimensional periodic structure of periodicity $p$ and composed of arbitrarily-shaped unit particles (e.g., spheres), as shown in Fig.~\ref{fig:gperiodic}. 

The driven-mode analysis procedure excites a finite ($N$-cell) version of the periodic structure at one of its ends with a harmonic wave of a given real temporal frequency, or frequency, $\omega_\text{r}$. As a result, the wave propagates (TW-regime) along the structure at this frequency and progressively loses energy due to conduction and/or radiation losses. The fields phasors have therefore the \emph{spatially} damped-harmonic waveform
\begin{equation}\label{eq:phasor}
{\psi}(x)=\text{e}^{-\gamma x},
\end{equation}
where $\gamma=\alpha+j\beta$, corresponding to~\eqref{eq:all_k} and~\eqref{eq:w_re}, is the complex spatial frequency or propagation constant of the structure, with $\alpha$ and $\beta$ being the attenuation and phase constants. Assuming the time-harmonic dependence $\text{e}^{+j\omega{t}}$, $\alpha$ positive (negative) corresponds to a decaying (growing) wave in a lossy (active) medium,  while $\beta$ positive (negative) corresponds to forward (backward) wave along $\hat{x}$. For example, for the forward propagating wave, Eq.~\eqref{eq:phasor} leads to the spacetime-dependent waveform $\psi(x,t)=\Re\{\psi(x)\text{e}^{j\omega_\text{r} t}\}=\text{e}^{-\alpha x}\cos\left(\omega_\text{r} t-\beta x\right)$. Due to the periodic nature of the problem, the fields are generally superpositions of Floquet space harmonics\footnote{The phase constant of the $n^\text{th}$ harmonic is $\beta_n(\omega_\text{r})=\beta_0(\omega_\text{r})+2\pi{n}/p$ where $\beta_0$ is the phase constant of the zeroth-order mode.} and sometimes, as in some LWAs, only one of these space harmonics plays a significant role, so that the structure behaves as a uniform (non-periodic) waveguide.

On the other hand, the eigen-mode analysis procedure specifies a phase difference, $\phi$, between the periodic boundaries of the structure (see Fig.~\ref{fig:gperiodic}), i.e., at the edges of the unit cell, which corresponds to the real spatial frequency $\beta=-\phi/p$. As a result, the wave resonates (SW-regime) along the unit cell, and the field phasors follow therefore the \emph{temporally} damped-harmonic waveform
\begin{equation}\label{eq:phasor_damaped_cosine}
\psi(t)=\text{e}^{j\omega{t}},
\end{equation}
where $\omega=\omega_\text{r}+j\omega_\text{i}$, corresponding to~\eqref{eq:all_w} and~\eqref{eq:k_re}, is the complex temporal frequency of the structure, with $\omega_\text{r}$ and $\omega_\text{i}$ being the usual frequency and the inverse of the relaxation time, as mentioned in Sec.~\ref{sec:intro}. Assuming the space-harmonic dependence $\text{e}^{-j\beta{x}}$, $\omega_\text{r}$ positive (negative) corresponds to forward (backward) wave along $\hat{x}$, while $\omega_\text{i}$ positive (negative) correspond to a decaying (growing) wave in a lossy (gain) medium. For example, for the forward propagating wave, Eq.~\eqref{eq:phasor_damaped_cosine} leads to the spacetime-dependent waveform $\psi(x,t)=\Re\{\psi(t)\text{e}^{-j\beta x}\}=\text{e}^{-\omega_\text{i} t}\cos\left(\omega_\text{r} t-\beta x\right)$.

\subsection{Complex Plane Mapping}\label{sec:general_structure}
As we have just seen, the driven-mode analysis gives the complex propagation constant $\gamma=\alpha+j\beta$ in terms of the purely real frequency $\omega=\omega_\text{r}$ [Eq.~
\eqref{eq:w_re}], while the eigen-mode analysis gives the complex frequency $\omega=\omega_\text{r}+j\omega_\text{i}$ in terms of the purely imaginary propagation constant $\gamma=j\beta$ [Eq.~\eqref{eq:k_re}]. 

The problem to solve is to find a mapping procedure that systematically relates the driven-mode (or TW-regime) problem and the eigen-mode (or SW/PBC-regime) problems, i.e., to transform the triplets ($\omega_\text{r}$, $\beta$, $\alpha$) into the triplet ($\beta$, $\omega_\text{r}$, $\omega_\text{i}$), and vice-versa.

\begin{figure}[h!]
	\centering
	\includegraphics[width=\columnwidth]{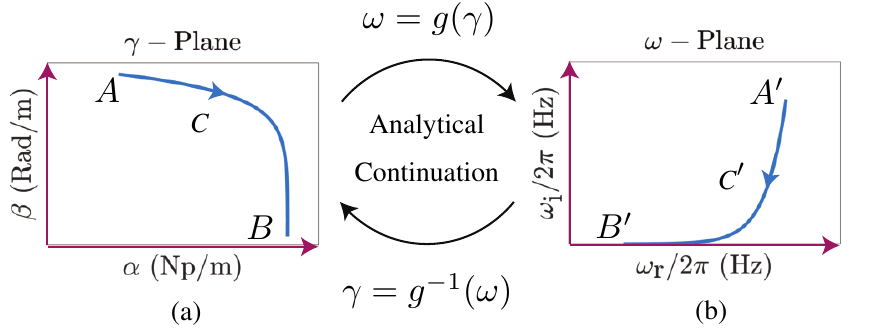}
	\caption{Statement of the general mapping problem: find, for any electromagnetic structure (also without analytical solution), the complex function $\omega=g(\gamma)$ that maps complex points in the $\gamma$-plane to complex points in the $\omega$-plane, and its reciprocal counterpart $\gamma=g^{-1}(\omega)$ that maps complex points in the $\omega$-plane to complex points in the $\gamma$-plane. Here, the complex function $\omega=g(\gamma)$ maps the points along the path $C$ between the points $A$ and $B$ in the $\gamma$-plane to the points along the path $C'$ between the points $A'$ and $B'$ in the $\omega$-plane while the complex function $\gamma=g^{-1}(\omega)$ performs the reverse mapping. The function $g(\cdot)$ will be shown to be unique, by virtue of the combined principles of analytical continuity and wave physicality (Sec.~\ref{sec:General_solution}).}\label{fig:gmapping}
\end{figure}

In fact, our solution to this problem will be even more generally map any pairs of complex regions of the spatial-frequency and temporal-frequency complex planes, by setting up a function $g$, which will turn out to be unique, such that $\omega=g(\gamma)$, mapping the complex $\gamma$-plane to the complex $\omega$-plane and, reciprocally, $\gamma=g^{-1}(\omega)$, mapping the complex $\omega$-plane to complex $\gamma$-plane, as illustrated in Fig.~\ref{fig:gmapping}.

The procedure should \emph{not} depend on the specific structure of interest and should be applicable to \emph{any} periodic or non-periodic structure.  

\section{Problems with Analytical Solutions}\label{sec:spec_examples}

We consider here three problems having a known analytical solution, both to further precise the problematic of the mapping problem introduced in Sec.~\ref{sec:prob_stat} and to establish analytical benchmarks for the validation in Sec.~\ref{sec:val_ill} of our general method, that will be presented in Sec.~\ref{sec:General_solution}.

\subsection{Lossy Medium}\label{sec:lossy_med}
Let us first assume the relatively simple problem of an unbounded lossy medium~\cite{dyab2015interpretation}. The wave functions of this problem, for instance $\psi=\mathcal{E}_z(x,t)$, satisfy the Helmholtz wave equation
\begin{equation}\label{eq:wave_equation}
\frac{\partial^2 \psi}{\partial t^2}-\frac{1}{\mu \epsilon}\frac{\partial^2 \psi}{\partial x^2}+\frac{\sigma}{\epsilon}\frac{\partial \psi}{\partial t}=0,
\end{equation}
where $\epsilon$, $\mu$ and $\sigma$ are respectively the permittivity, the permeability and the conductivity of the medium. A solution to~\eqref{eq:wave_equation} is 
\begin{align}\label{eq:plane_wave_space}
\mathcal{E}_z(x,t)&=\text{e}^{-\alpha x}\cos({\omega_\text{r} t-\beta x)} \nonumber \\
&=\Re\{\text{e}^{-\gamma x}\text{e}^{j\omega_\text{r} t}\},
\end{align}
where $\gamma=\alpha+j\beta$ is the complex propagation constant. Substituting \eqref{eq:plane_wave_space} in \eqref{eq:wave_equation} results into
\begin{equation}\label{eq:propagation_constant}
\gamma(\omega_\text{r})=\pm j\frac{\omega_\text{r}}{\nu}\sqrt{1-j\frac{\sigma}{\omega_\text{r}\epsilon}},
\end{equation}
where $\nu=1/\sqrt{\mu \epsilon}$ is the speed of the wave in the medium. Equation~\eqref{eq:propagation_constant} provides the complex propagation constant, $\gamma$, corresponding to the real frequency, $\omega_\text{r}$, for the medium parameters $\epsilon$, $\mu$ and $\sigma$.

Alternatively, fixing the magnitude of the wave in space, i.e., setting $\alpha=0$, and letting the phase vary, i.e., $\beta \neq 0$, leads to
\begin{align}\label{eq:plane_wave_time}
\mathcal{E}_z(x,t)&=\Re\{\text{e}^{-j\beta x}\text{e}^{j\omega t}\} \nonumber \\
&=\text{e}^{-\omega_\text{i} t}\cos({\omega_\text{r} t-\beta x)}, 
\end{align}
with complex frequency $\omega=\omega_\text{r}+j\omega_\text{i}$, which is also solution to~\eqref{eq:wave_equation}. Substituting \eqref{eq:plane_wave_time} in \eqref{eq:wave_equation} results into
\begin{equation}\label{eq:phase_constant}
\beta=\pm \frac{\omega}{\nu}\sqrt{1-j\frac{\sigma}{\omega\epsilon}},
\end{equation}    
which is exactly Eq.~\eqref{eq:propagation_constant} except that $\gamma$ and $\omega_\text{r}$ are replaced by $j\beta$ and $\omega$, respectively. Solving Eq.~\eqref{eq:phase_constant} for $\omega$ yields
\begin{equation}\label{eq:complex_frequency}
\omega(\beta)=\pm\sqrt{\nu^2\beta^2-\frac{\sigma^2}{4\epsilon^2}}+j\frac{\sigma}{2\epsilon}.
\end{equation} 
Equation~\eqref{eq:complex_frequency} provides the complex frequency $\omega$, given the phase constant $\beta$ and the medium parameters $\epsilon$, $\mu$ and $\sigma$. 

\begin{figure}[h!]
	\begin{center}
		\subfigure[] {
			\includegraphics[width=0.46\columnwidth]{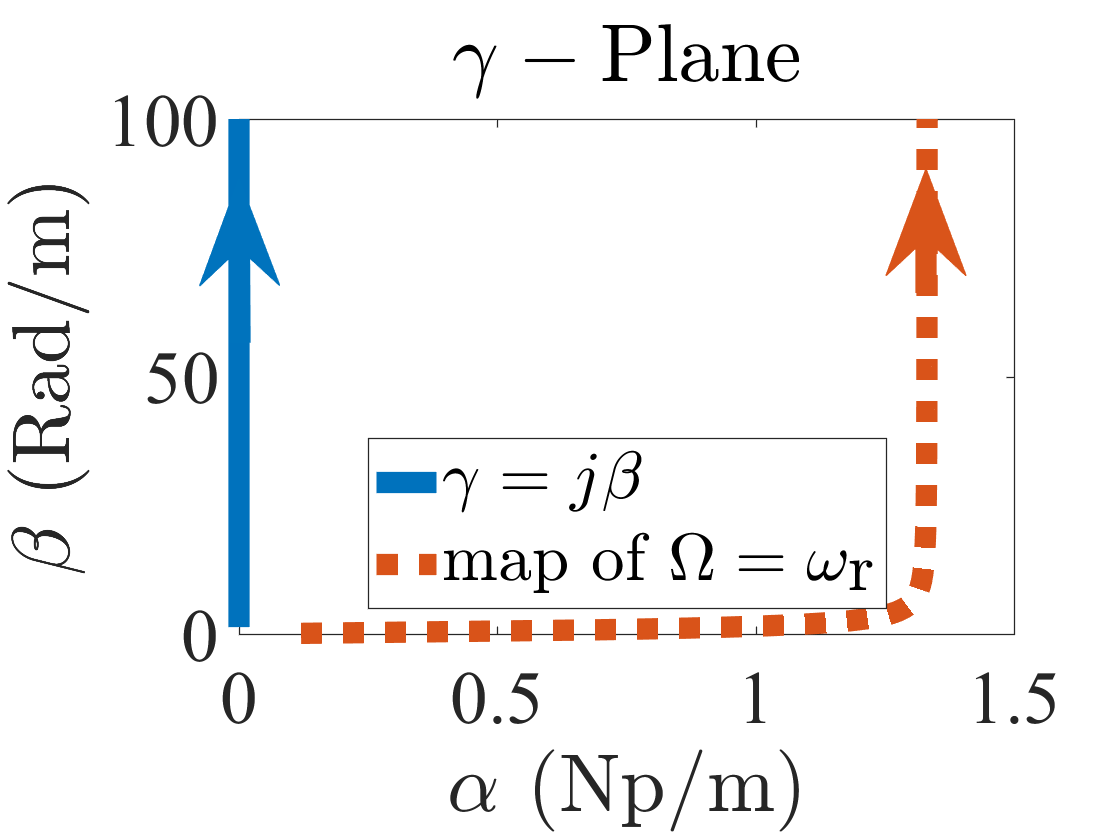}
		}
		\subfigure[] {
			\includegraphics[width=0.46\columnwidth]{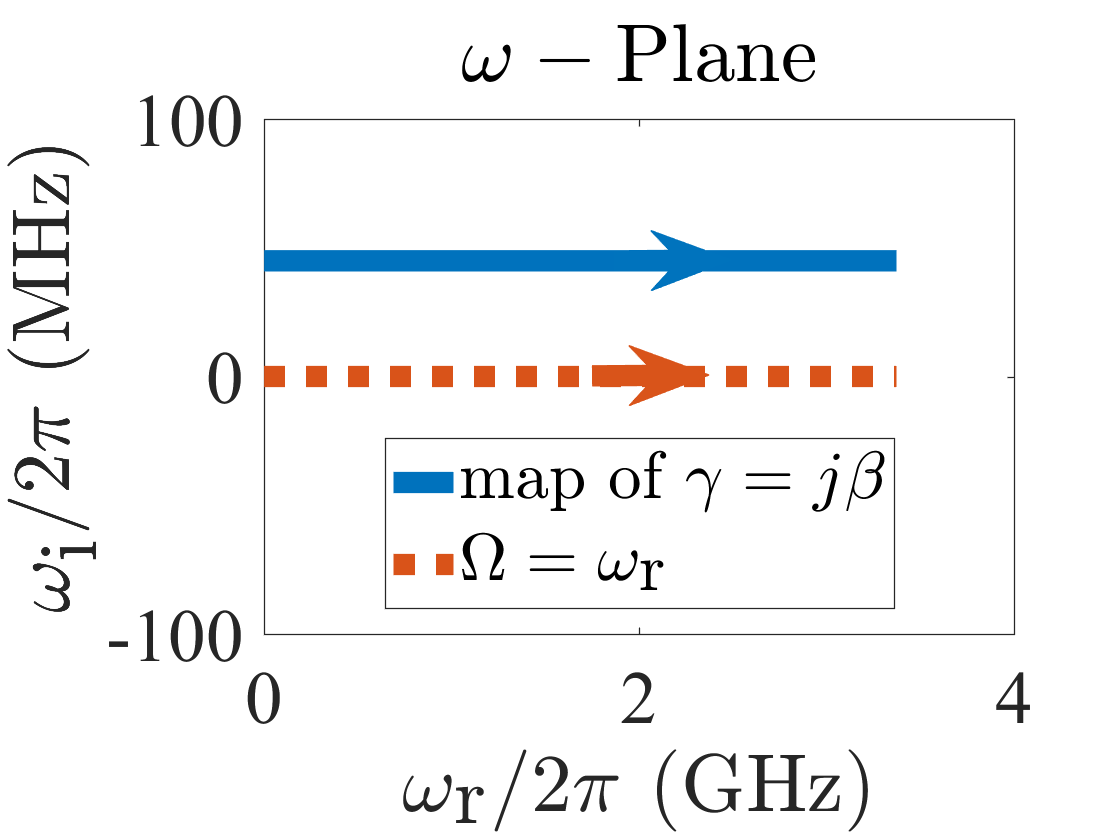}
		}
		\caption{Mapping for an unbounded lossy medium with relative permittivity $\epsilon_\text{r}=2.2$ and conductivity $\sigma=0.01~\textrm{S/m}$. (a)~$\gamma$-plane, with $\omega=\omega_\text{r}$ mapping using Eq.~\eqref{eq:propagation_constant}. (b) $\omega$-plane, with $\gamma=j\beta$ mapping using Eq.~\eqref{eq:complex_frequency}.}\label{Fig:gamma_omega_plane_UM}
	\end{center}
\end{figure}

Figure~\ref{Fig:gamma_omega_plane_UM}(a) shows $\gamma=j\beta$ in the  $\gamma$-plane while Fig.~\ref{Fig:gamma_omega_plane_UM}(b) shows its map in the $\omega$-plane given by the Eq.~\eqref{eq:complex_frequency}. Conversely, Fig.~\ref{Fig:gamma_omega_plane_UM}(b) shows the $\omega=\omega_\text{r}$ in the $\omega$-plane while Fig.~\ref{Fig:gamma_omega_plane_UM}(a) shows its map in the $\gamma$-plane given by the Eq.~\eqref{eq:propagation_constant}.   

A couple of important observations are in order. 
\begin{enumerate}
	\item According to Eq.~\eqref{eq:complex_frequency}, the real part of the complex frequency $\omega_\text{r}=\pm \sqrt{\nu^2\beta^2-{\sigma^2}/{4\epsilon^2}}$ is approximately $\omega_\text{r} \approx \pm \nu\beta$\footnote{We usually choose the positive sign for the real frequency $\omega_\text{r}=\sqrt{\nu^2\beta^2-{\sigma^2}/{4\epsilon^2}} \approx \nu\beta$.} in a low-loss dielectric medium, where $\sigma/2\epsilon \ll \nu\beta$, and the imaginary part of $\omega$ is then a positive constant $\omega_\text{i}={\sigma}/{2\epsilon}$\footnote{According to Eq.~\eqref{eq:plane_wave_space}, the fields damps as it propagates in space. Likewise, according to Eq.~\eqref{eq:plane_wave_time}, the wave damps in time.}. 
	\item According to Eq.~\eqref{eq:propagation_constant} and choosing the positive sign for the forward propagating waves, the real part of $\gamma$ is approximately constant $\alpha \approx \sigma/2\epsilon\nu$ and the imaginary part of $\gamma$ is approximately $\beta \approx \omega_\text{r}/\nu$ for a low loss medium where $\sigma/\omega_\text{r}\epsilon \ll 1$.  
\end{enumerate}
 
The equations~\eqref{eq:plane_wave_space} and~\eqref{eq:plane_wave_time} are both solutions of the Helmholtz equation~\eqref{eq:wave_equation}, but have different boundary conditions. Equation~\eqref{eq:plane_wave_space} is the solution for the driven-mode regime while Eq.~\eqref{eq:plane_wave_time} is the solution for the eigen-mode regime. If the eigen-mode regime would accept magnitude differences at the boundaries in addition to the phase differences, then the general representation
\begin{equation}\label{eq:plane_wave_genral}
\psi(x,t)=\Re\{\text{e}^{-\gamma x}\text{e}^{j\omega t}\}, 
\end{equation}  
would lead to a solution of the wave equation~\eqref{eq:wave_equation} with $\gamma$ and $\omega$ both complex.     

Inserting Eq.~\eqref{eq:plane_wave_genral} into~\eqref{eq:wave_equation} results in the complex dispersion relation
\begin{equation}\label{eq:relation_Omega_gamma_UBM}
\omega^2+\frac{1}{\mu \epsilon}\gamma^2-j\frac{\sigma}{\epsilon}\omega=0,
\end{equation} 
which reduces to Eq.~\eqref{eq:propagation_constant} if $\omega=\omega_\text{r}$ and to Eq.~\eqref{eq:complex_frequency} if $\gamma=j\beta$. Equation~\eqref{eq:relation_Omega_gamma_UBM} provides the most general mapping function $g$ from the $\gamma$-plane into the $\omega$-plane, given by
\begin{subequations}
\begin{align}\label{eq:complex_frequency_gamma_UM}
\omega&=g(\gamma) \nonumber \\
&=j\left(\frac{\sigma}{2\epsilon} \pm \sqrt{\nu^2\gamma^2+\frac{\sigma^2}{4\epsilon^2}}\right),
\end{align} 
while, inversely, the mapping function $g^{-1}$ from the $\omega$-plane into the $\gamma$-plane, given by
\begin{align}\label{eq:complex_gamma_frequency_UM}
\gamma&=g^{-1}(\omega) \nonumber \\
&=\pm j\frac{\omega}{\nu}\sqrt{1-j\frac{\sigma}{\omega\epsilon}}.
\end{align}  
\end{subequations}
 
\subsection{Dielectric-filled Metallic Waveguide}\label{sec:met_wg}
Second, we consider the rectangular metallic waveguide filled with a lossy dielectric material~\cite{king2019relation}. The wave functions of this problem, for instance $\psi=\mathcal{H}_x(x,t)$ or $\psi=\mathcal{E}_x(x,t)$, satisfy the Helmholtz wave equation
\begin{equation}\label{eq:wave_equation_WG}
{\mu \epsilon}\frac{\partial^2 \psi}{\partial t^2}-\frac{\partial^2 \psi}{\partial x^2}+\kappa_{m,n}^2\psi+{\mu\sigma}\frac{\partial \psi}{\partial t}=0,
\end{equation}
where $\kappa_{m,n}=\sqrt{\left({m\pi}/{a}\right)^2+\left({n\pi}/{b}\right)^2}$, with $a$ and $b$ representing the dimensions of the rectangular cross section of the waveguide, and ($m$,$n$) denotes the mode number.  

Inserting Eq.~\eqref{eq:plane_wave_space} in \eqref{eq:wave_equation_WG} yields
\begin{equation}\label{eq:propagation_constant_WG}
\gamma(\omega_\text{r})=\pm j\frac{\omega_\text{r}}{\nu}\sqrt{1-\left(\frac{\nu\kappa_{m,n}}{\omega_\text{r}}\right)^2-j\frac{\sigma}{\omega_\text{r}\epsilon}},
\end{equation}
while inserting Eq.~\eqref{eq:plane_wave_time} in \eqref{eq:wave_equation_WG} yields
\begin{equation}\label{eq:complex_frequency_WG}
\omega(\beta)=\pm\sqrt{\nu^2\left(\beta^2+\kappa_{m,n}^2\right)-\frac{\sigma^2}{4\epsilon^2}}+j\frac{\sigma}{2\epsilon}.
\end{equation}

\begin{figure}[h!]
	\begin{center}
		\subfigure[] {
			\includegraphics[width=0.46\columnwidth]{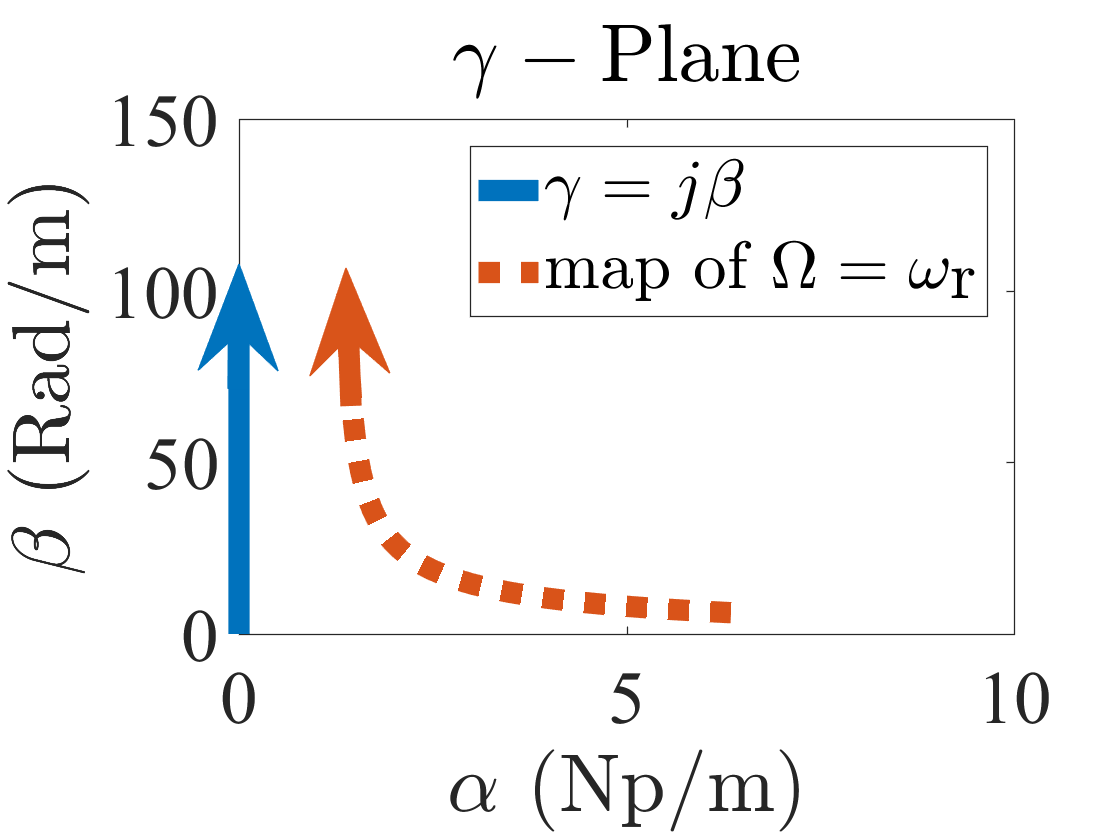}
		}
		\subfigure[] {
			\includegraphics[width=0.46\columnwidth]{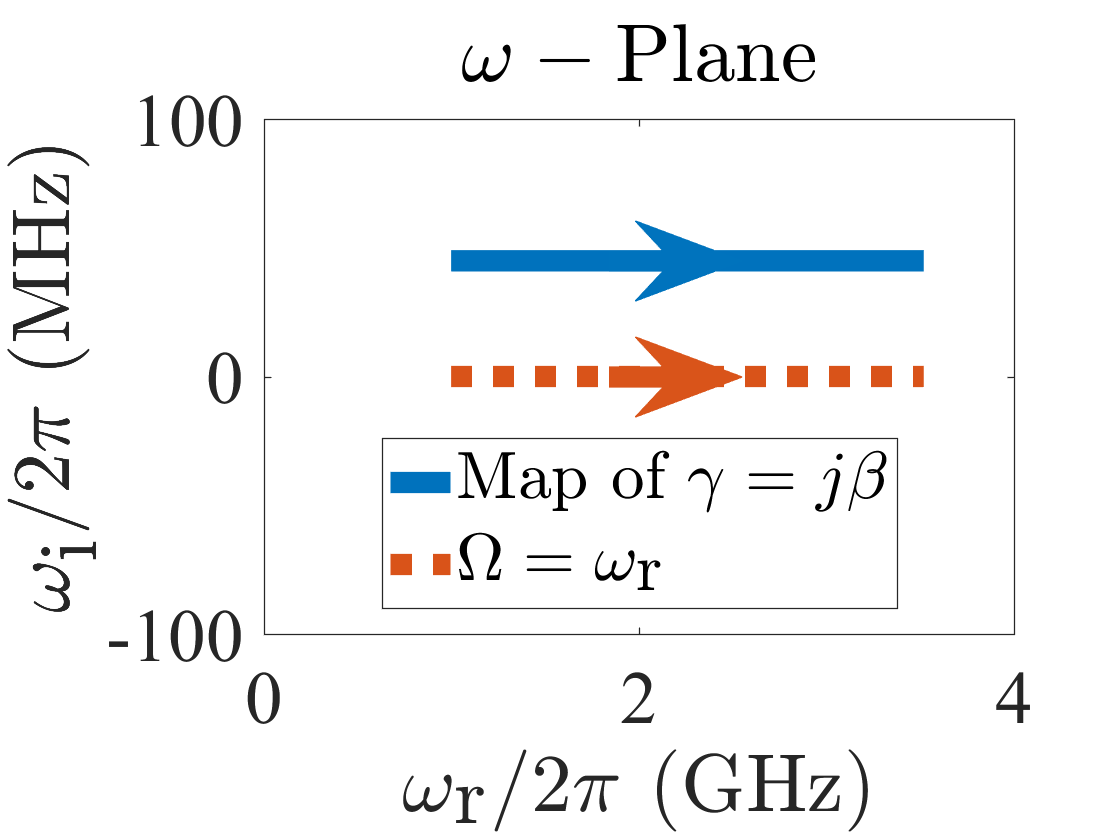}
		}
		\caption{Mapping for a rectangular metallic waveguide filled with a lossy dielectric material with $\epsilon_\text{r}=2.2$ and $\sigma=0.01~\textrm{S/m}$ of width $a=\lambda_\text{g}/2$ where $\lambda_\textrm{g}=\lambda_0/\sqrt{\epsilon_\text{r}}$ is the wavelength in the dielectric medium and $\lambda_0=30~\text{cm}$, corresponding to the cut-off frequency of 1~GHz for the $\text{TE}_{10}$ mode. (a)~$\gamma$-plane, with $\omega=\omega_\text{r}$ mapping using Eq.~\eqref{eq:propagation_constant_WG}. (b) $\omega$-plane, with $\gamma=j\beta$ mapping using Eq.~\eqref{eq:complex_frequency_WG}.}
		\label{Fig:gamma_omega_plane_WG}
	\end{center}
\end{figure}

Figure~\ref{Fig:gamma_omega_plane_WG}(a) shows $\gamma=j\beta$ in the $\gamma$-plane while Fig.~\ref{Fig:gamma_omega_plane_WG}(b) shows its map in the $\omega$-plane given by the Eq.~\eqref{eq:complex_frequency_WG} and reversely, Fig.~\ref{Fig:gamma_omega_plane_WG}(b) shows the $\omega=\omega_\text{r}$ in the $\omega$-plane and Fig.~\ref{Fig:gamma_omega_plane_WG}(a) shows its map in the $\gamma$-plane given by the Eq.~\eqref{eq:propagation_constant_WG} for the $\text{TE}_{10}$ mode where $m=1$ and $n=0$ (i.e., $\kappa_{m,n}=\pi/a$).

A couple of important observations are in order. 
\begin{enumerate}
	\item According to Eq.~\eqref{eq:complex_frequency_WG}, the real part of the complex frequency $\omega_\text{r}=\pm \sqrt{\nu^2\tilde{\beta}^2-{\sigma^2}/{4\epsilon^2}}$ (where $\tilde{\beta}^2={\beta^2+\kappa_{m,n}^2}$) is approximately $\omega_\text{r} \approx \pm \nu\tilde{\beta}$ for a low-loss medium where $\sigma/2\epsilon \ll \nu \tilde{\beta}$, and the imaginary part of $\omega$ is then a positive constant $\omega_\text{i}={\sigma}/{2\epsilon}$. 
	\item According to Eq.~\eqref{eq:propagation_constant_WG}, and choosing the positive sign for the forward propagating waves, the real part of $\gamma$ is approximately $\alpha \approx (\omega_\text{r}/\tilde{\omega_\text{r}})\sigma/2\epsilon\nu$ (where $\tilde{\omega_\text{r}}=\sqrt{\omega_\text{r}^2-\nu^2\kappa_{m,n}^2}$) and the imaginary part of $\gamma$ is approximately $\beta \approx \tilde{\omega_\text{r}}/\nu$ for a low-loss medium where $\omega_\text{r}\sigma/\tilde{\omega_\text{r}}^2\epsilon \ll 1$.  
\end{enumerate}

Equations \eqref{eq:propagation_constant_WG} and \eqref{eq:complex_frequency_WG} are solutions of the wave equation~\eqref{eq:wave_equation_WG}. Inserting the general wave representation~\eqref{eq:plane_wave_genral} into~\eqref{eq:wave_equation_WG}, we obtain the following general dispersion equation    
\begin{equation}\label{eq:relation_Omega_gamma_WG}
\omega^2+\frac{1}{\mu \epsilon}(\gamma^2-\kappa_{m,n}^2)-j\frac{\sigma}{\epsilon}\omega=0,
\end{equation}  
which reduces to Eq.~\eqref{eq:propagation_constant_WG} if $\omega=\omega_\text{r}$ and to Eq.~\eqref{eq:complex_frequency_WG} if $\gamma=j\beta$. Equation~\eqref{eq:relation_Omega_gamma_WG}
results in the mapping function $g$ from the $\gamma$-plane into the $\omega$-plane, given by
\begin{subequations}
\begin{align}\label{eq:complex_frequency_gamma_WG}
\omega&=g(\gamma) \nonumber \\
&=j\left(\frac{\sigma}{2\epsilon} \pm \sqrt{\nu^2(\gamma^2-\kappa_{m,n}^2)+\frac{\sigma^2}{4\epsilon^2}}\right),
\end{align} 
and inversely the mapping function $g^{-1}$ from the $\omega$-plane into the $\gamma$-plane, given by
\begin{align}\label{eq:complex_gamma_frequency_WG}
\gamma&=g^{-1}(\omega) \nonumber \\
&=\pm j\frac{\omega}{\nu}\sqrt{1-\left(\frac{\nu\kappa_{m,n}}{\omega}\right)^2-j\frac{\sigma}{\omega\epsilon}}.
\end{align} 
\end{subequations}

\subsection{Periodically-loaded Transmission Line}\label{sec:periodic_loaded_PLTL}
Finally, let us consider a loaded transverse electromagnetic (TEM) transmission line (TL) that is periodically loaded by shunt lossy capacitive loads with periodicity $p$~\cite{pozar2011microwave}. 

The voltage or current functions, $\psi=\mathcal{V}(x,t)$ and $\psi=\mathcal{I}(x,t)$, of such a TL satisfy the wave equation
\begin{equation}\label{eq:wave_equation_TL}
{L C}\frac{\partial^2 \psi}{\partial t^2}-\frac{\partial^2 \psi}{\partial x^2}+RG\psi+\left(L G + R C\right)\frac{\partial \psi}{\partial t}=0,
\end{equation}  
where $R$ ($\omega$/m), $G$ (S/m), $L$ (H/m) and $C$ (F/m) are the per-unit-length series resistance, shunt conductance, series inductance and shunt capacitance, respectively. On the other hand, the voltage $V(t)$ and current $I(t)$ for a shunt lossy capacitive load satisfy the equation
\begin{equation}\label{eq:equation_load}
\tilde{C}\frac{d V}{d t}+\tilde{G}{V}=I,
\end{equation}   
where $\tilde{C}$ (F) and $\tilde{G}$ (S) are capacitance and conductance of the lumped load, respectively. 
Equations~\eqref{eq:wave_equation_TL} and \eqref{eq:equation_load} both admit solutions of the form \eqref{eq:plane_wave_space}, \eqref{eq:plane_wave_time} and \eqref{eq:plane_wave_genral}. 

The relation between the voltages and currents at either side of the $n^{\text{th}}$ unit cell is given by~\cite{pozar2011microwave}  
\begin{equation}\label{eq:ABCD}
\begin{pmatrix} V_n \\ I_n \end{pmatrix}=
\begin{pmatrix} A&B \\ C&D \end{pmatrix} 
\begin{pmatrix} V_{n+1} \\ I_{n+1} \end{pmatrix},
\end{equation}
where $A$, $B$, $C$ and $D$ are the transmission matrix parameters of a cascaded load-free $p/2$-length TL, shunt admittance $Y$ and $p/2$-length unloaded TL. Equating the $ABCD$ matrix to that of a section of an equivalent TL with length $p$ and complex propagation constant $\gamma$ leads to (see~\cite{pozar2011microwave}, Sec. 8, page 383)  
\begin{equation}\label{eq:complex_gamma_PLTL}
\gamma(\omega_\text{r})=\frac{1}{p}\cosh^{-1}\left[ \cos \left(\frac{\omega_\text{r}}{\nu}p\right) -\frac{b}{2}\sin \left(\frac{\omega_\text{r}}{\nu}p\right) \right],
\end{equation}
where $b=-jYZ_0$ and $Z_0$ is the characteristic impedance of the unloaded TL. The function $\gamma(\omega_\text{r})$ is periodic with periodicity $\nu\pi/p$ according to Eq.~\eqref{eq:complex_gamma_PLTL}. 

On the other hand, if we replace $\omega_\text{r}$ by the complex $\omega$ and the complex $\gamma$ by purely imaginary $j\beta$ in Eq.~\eqref{eq:complex_gamma_PLTL}, we find the complex $\omega$ given by 
\begin{subequations}\label{eq:complex_frequency_PLTL}
\begin{equation}
\omega(\beta)=\frac{\nu}{p} \left(\cos^{-1}\left[\cos(\beta p) \cos \phi\right] - \phi \right),
\end{equation}
where
\begin{equation}
\phi=\tan^{-1}\left(\frac{b}{2}\right).
\end{equation}
\end{subequations}
The function $\omega(\beta)$ is periodic with periodicity $\pi/p$ according to Eq.~\eqref{eq:complex_frequency_PLTL}.
\begin{figure}[h!]
	\begin{center}
		\subfigure[] {
			\includegraphics[width=0.46\columnwidth]{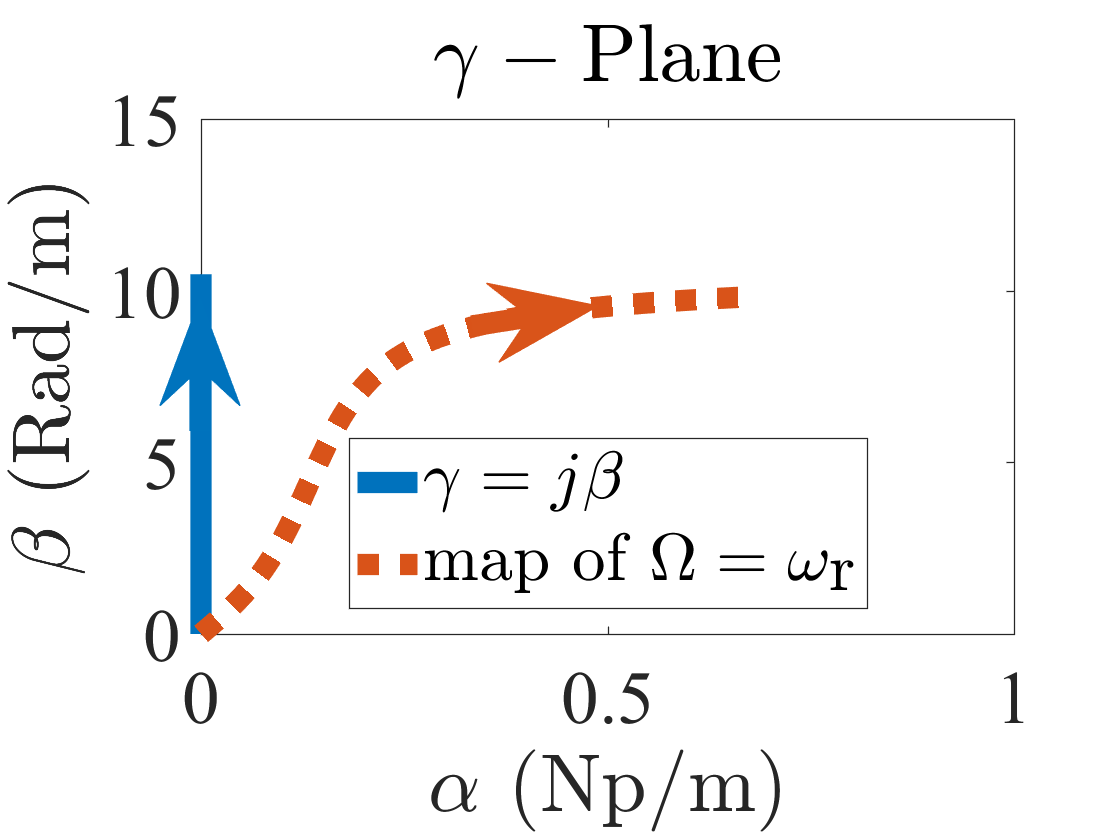}
		}
		\subfigure[] {
			\includegraphics[width=0.46\columnwidth]{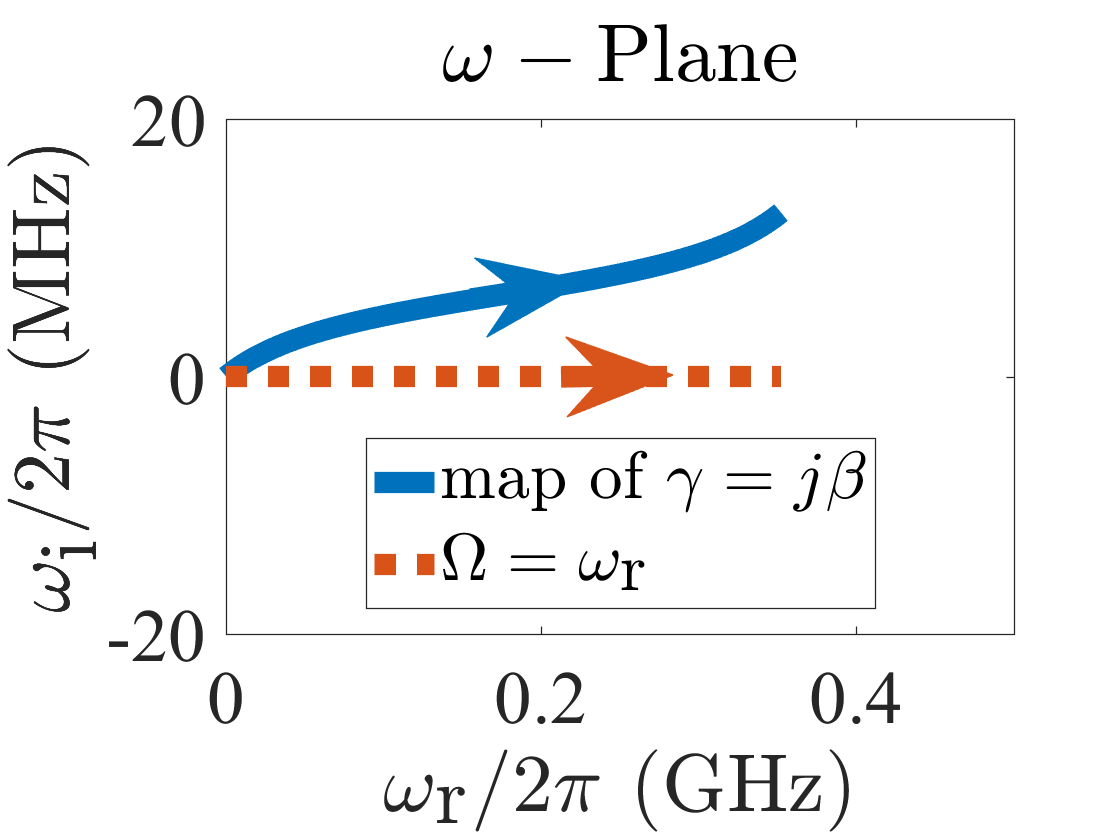}
		}
		\caption{Mapping for a TL periodically loaded by lossy capacitive loads with normalized admittance $YZ_0 = 0.1+j$ and periodicity $p=30$ cm. (a)~$\gamma$-plane, with $\omega=\omega_\text{r}$ mapping using Eq.~\eqref{eq:complex_gamma_PLTL}. (b) $\omega$-plane, with $\gamma=j\beta$ mapping using Eq.~\eqref{eq:complex_frequency_PLTL}. $\gamma$ is periodic with periodicity $j\pi/p \approx j10.5~(\text{Rad}/\text{m})$ and $\omega/2\pi$ is periodic with periodicity $\nu/2p \approx 0.5$ GHz.}\label{Fig:gamma_omega_plane_PLTL}
	\end{center}
\end{figure}

Letting $\omega$ and $\gamma$ be both complex in Eq.~\eqref{eq:complex_gamma_PLTL}, we obtain the general dispersion equation    
\begin{equation}\label{eq:relation_Omega_gamma_PLTL}
\cosh\gamma p=\cos\left(\frac{\omega}{\nu}p\right) -\frac{b}{2}\sin\left(\frac{\omega}{\nu}p\right), 
\end{equation}  
which reduces to Eq.~\eqref{eq:complex_gamma_PLTL} if $\omega=\omega_\text{r}$ and to Eq.~\eqref{eq:complex_frequency_PLTL} if $\gamma=j\beta$.
Equation~\eqref{eq:relation_Omega_gamma_PLTL}
provides the mapping function $g$ from the $\gamma$-plane into the $\omega$-plane,
\begin{subequations}
\begin{align}\label{eq:complex_frequency_gamma_PLTL}
\omega&=g(\gamma) \nonumber \\
&=\frac{\nu}{p} \left(\cos^{-1}\left[\cosh(\gamma p) \cos \phi\right] - \phi \right).
\end{align} 
and inversely the mapping function $g^{-1}$ from the $\omega$-plane into the $\gamma$-plane, 
\begin{align}\label{eq:complex_gamma_frequency_PLTL}
\gamma&=g^{-1}(\omega) \nonumber \\
&=\frac{1}{p}\cosh^{-1}\left[ \cos \left(\frac{\omega}{\nu}p \right) -\frac{b}{2}\sin \left(\frac{\omega}{\nu}p \right) \right].
\end{align}
\end{subequations}

\section{General Mapping using Analytic Continuation}\label{sec:General_solution}
\subsection{Motivation}\label{sec:motivation}
Section~\ref{sec:spec_examples} described the complex frequency mapping of some canonical problems that admit analytical solutions, namely an unbounded lossy medium, a dielectric-filled rectangular WG and a periodically-loaded TL. In these cases, we could obtain \emph{closed-form expressions} for the mapping functions -- $\gamma(\omega_\text{r})$, $\omega(\beta)$, $\omega=g(\gamma)$ and $\gamma=g^{-1}(\omega)$ -- from the wave equations corresponding to the structure, specifically from Eqs.~\eqref{eq:wave_equation},~\eqref{eq:wave_equation_WG}\footnote{For the waveguide problem, we  need specify a priori the mode number ($m$,$n$) to find $\omega=g(\gamma)$.} and~\eqref{eq:wave_equation_TL} given the simplicity of the boundary conditions. Table~\ref{tab:examples_analytic_sol} summarizes the mapping functions obtained for these problems with reference to the equation numbers in the text.
\begin{table*}[ht!]
	\caption[caption]{Mapping functions $\gamma(\omega_\text{r})$, $\omega(\beta)$, $\omega=g(\gamma)$ and $\gamma=g^{-1}(\omega)$ for the 3 examples in Sec.~\ref{sec:spec_examples}.}\label{tab:examples_analytic_sol}
	\normalsize
	\centering
	\setlength{\tabcolsep}{5pt}
	\begin{tabular}{c|c c c}
		\hline \\[-0.8em]
		\text{Examples} &
		\text{lossy medium} &
		\text{rectangular WG} &
		\text{periodically-loaded TL}\\
		\hline \\[-0.6em]
		$\gamma(\omega_\text{r})$&
		$\pm j\frac{\omega_\text{r}}{\nu}\sqrt{1-j\frac{\sigma}{\omega_\text{r}\epsilon}}$&
		$\pm j\frac{\omega_\text{r}}{\nu}\sqrt{1-\left(\frac{\nu\kappa_{m,n}}{\omega_\text{r}}\right)^2-j\frac{\sigma}{\omega_\text{r}\epsilon}}$&
		$\frac{1}{p}\cosh^{-1}\left[ \cos \left(\frac{\omega_\text{r}}{\nu}p\right) -\frac{b}{2}\sin \left(\frac{\omega_\text{r}}{\nu}p\right) \right]$
		\\ 
		&
		Eq.~\eqref{eq:propagation_constant}&
	    Eq.~\eqref{eq:propagation_constant_WG}&
		Eq.~\eqref{eq:complex_gamma_PLTL} 
		\\ [0.6em]
		\hline
		\\ [-0.6em]		
		$\omega(\beta)$&
		$\pm\sqrt{\nu^2\beta^2-\frac{\sigma^2}{4\epsilon^2}}+j\frac{\sigma}{2\epsilon}$&
		$\pm\sqrt{\nu^2\left(\beta^2+\kappa_{m,n}^2\right)-\frac{\sigma^2}{4\epsilon^2}}+j\frac{\sigma}{2\epsilon}$&
		$\frac{\nu}{p} \left(\cos^{-1}\left[\cos(\beta p) \cos \phi\right] - \phi \right)$ 
		\\ 
		&
		Eq.~\eqref{eq:complex_frequency}&
		Eq.~\eqref{eq:complex_frequency_WG}&
		Eq.~\eqref{eq:complex_frequency_PLTL}
		\\ [0.6em]
		\hline
		\\ [-0.6em]
		$\omega=g(\gamma)$&
		$j\left(\frac{\sigma}{2\epsilon} \pm \sqrt{\nu^2\gamma^2+\frac{\sigma^2}{4\epsilon^2}}\right)$&
		$j\left(\frac{\sigma}{2\epsilon} \pm \sqrt{\nu^2(\gamma^2-\kappa_{m,n}^2)+\frac{\sigma^2}{4\epsilon^2}}\right)$&
		$\frac{\nu}{p} \left(\cos^{-1}\left[\cosh(\gamma p) \cos \phi\right] - \phi \right)$ 
		\\ 
		&
		Eq.~\eqref{eq:complex_frequency_gamma_UM}&
		Eq.~\eqref{eq:complex_frequency_gamma_WG}&
		Eq.~\eqref{eq:complex_frequency_gamma_PLTL}
		\\ [0.6em]
		\hline
		\\ [-0.6em]
		$\gamma=g^{-1}(\omega)$&
		$\pm j\frac{\omega}{\nu}\sqrt{1-j\frac{\sigma}{\omega\epsilon}}$&
		$\pm j\frac{\omega}{\nu}\sqrt{1-\left(\frac{\nu\kappa_{m,n}}{\omega}\right)^2-j\frac{\sigma}{\omega\epsilon}}$&
		$\frac{1}{p}\cosh^{-1}\left[ \cos \left(\frac{\omega}{\nu}p \right) -\frac{b}{2}\sin \left(\frac{\omega}{\nu}p \right) \right]$ 
		\\ 
		&
		Eq.~\eqref{eq:complex_gamma_frequency_UM}&
		Eq.~\eqref{eq:complex_gamma_frequency_WG}&
		Eq.~\eqref{eq:complex_gamma_frequency_PLTL}
		\\ [0.6em]
		\hline
	\end{tabular}
	\vspace*{4pt}
\end{table*}

The availability of closed-form expressions for the mapping functions, as in Tab.~\ref{tab:examples_analytic_sol}, naturally represents an ideal situation, since such solutions are exact, insightful and straightforward. Unfortunately, most practical systems, such as for instance a well-designed LWA structures~\cite{otto2011transmission}, are too complex to admit closed-form, and even analytic, solutions. The analysis of such systems require full-wave simulations, which deliver \emph{purely numerical} solutions. An approach capable to handle such problems is clearly needed. 

\subsection{General Problem and Resolution Procedure}\label{sec:general_problem}
Specifically, a general method is needed to map the \emph{numerical} complex temporal frequencies of eigen-mode simulations into the corresponding complex spatial frequencies or, inversely, the \emph{numerical} complex spatial frequencies of driven-mode simulations into the corresponding complex temporal frequencies\footnote{Eigen-mode and driven-mode solvers are in available commercial packages such as Ansys HFSS.}. For this purpose, one has to find a \emph{generic function} $g(\cdot)$, such that $\omega=g(\gamma)$ along the line $\gamma=j\beta$ of the complex $\gamma$ plane (eigen-mode computation) and $\gamma=g^{-1}(\omega)$ along the line $\omega=\omega_\text{r}$ of the complex $\omega$ plane (driven-mode computation).  

We shall solve the problem by using the following four-step procedure:
\begin{enumerate}
	\item Run a full-wave eigen-mode simulation\footnote{In the eigen-mode solver, $\beta$ is related to the periodic phase $\phi$ as $\beta=-\phi/p$ (see Fig.~\ref{fig:gperiodic}).} or a driven-mode simulation\footnote{The procedure to extract the complex propagation constant of a periodic structure from the computed simulated scattering parameters is described in the fifth paragraph of Sec.~\ref{sec:intro}.} of the problem to address so as to compute the data $\omega(\beta)=\omega_\text{r}(\beta)+j\omega_\text{i}(\beta)$ or $\gamma(\omega_\text{r})=\alpha(\omega_\text{r})+j\beta(\omega_\text{r})$, and record the numerical results, $[\beta,\omega_\text{r},\omega_\text{i}]$ or $[\omega_\text{r},\beta,\alpha]$, respectively.
	\item Express the complex mapping function $g(\cdot)$ or $g^{-1}(\cdot)$ as polynomial expansion of $\beta$ around an analytic point $\beta_0$ or of $\omega_\text{r}$ around an analytic point $\omega_{\text{r},0}$, determine the expansion coefficients by fitting to the results of 1), and build the corresponding routine $g(\beta)$ or $g^{-1}(\omega_\text{r})$ in a calculator, respectively.
	\item Substitute $\beta\rightarrow-j\gamma$ into the routine $g(\beta)$ or  $\omega_\text{r}\rightarrow{j}\omega$ into the routine $g^{-1}(\omega_\text{r})$, respectively. 
	\item Find the complex roots $\gamma=\alpha+j\beta$ of the equation $\omega_\text{r}-g(\gamma)=0$ or $\omega=\omega_\text{r}+j\omega_\text{i}$ of the equation $j\beta-g^{-1}(\omega)=0$, respectively, which represent the sought after mapped solutions.
\end{enumerate}

At this point the nature of the `function' $g(\cdot)$ is still rather vague. However, we will be able to precise it shortly.

\subsection{Analycity of the Solution}\label{sec:anal_sol}
We are interested here in engineering problems, whose solutions obviously have a \emph{physical reality}. Since the mathematical complex functions that describe physical problems are necessarily \emph{analytic}~\cite{morse1953methods}, we can assert that the \emph{function $g(\cdot)$ must be analytic}. We can then leverage the properties of analytic functions to assist us in characterizing $g(\cdot)$.

\subsection{Theorem of Analytic Continuation}\label{sec:act}
The most useful property of analytic functions toward the resolution of our problem is the property of \emph{analytic continuation}. The theorem of analytic continuation states the following~\cite{brown2009complex} (Sec. 27, page 84): \emph{A function that is analytic in a domain \textbf{D} is uniquely determined over \textbf{D} by its values in that domain or along any line segment contained in it.}

Since we typically know the function only along line segments (Sec.~\ref{sec:general_problem}), rather than in complete domains of analycity, the second part of the theorem, seems particularly convenient to us. We shall therefore focus on the statement: \emph{A function that is analytic in a domain \textbf{D} is uniquely determined over \textbf{D} by its values along any line segment contained in it}. In other words, if an analytic function is known over a line segment of its analytic region, it is known  \emph{everywhere}, since this function is unique across the entire domain of analycity of the function.

\begin{figure}[ht!]
	\includegraphics[width=\columnwidth]{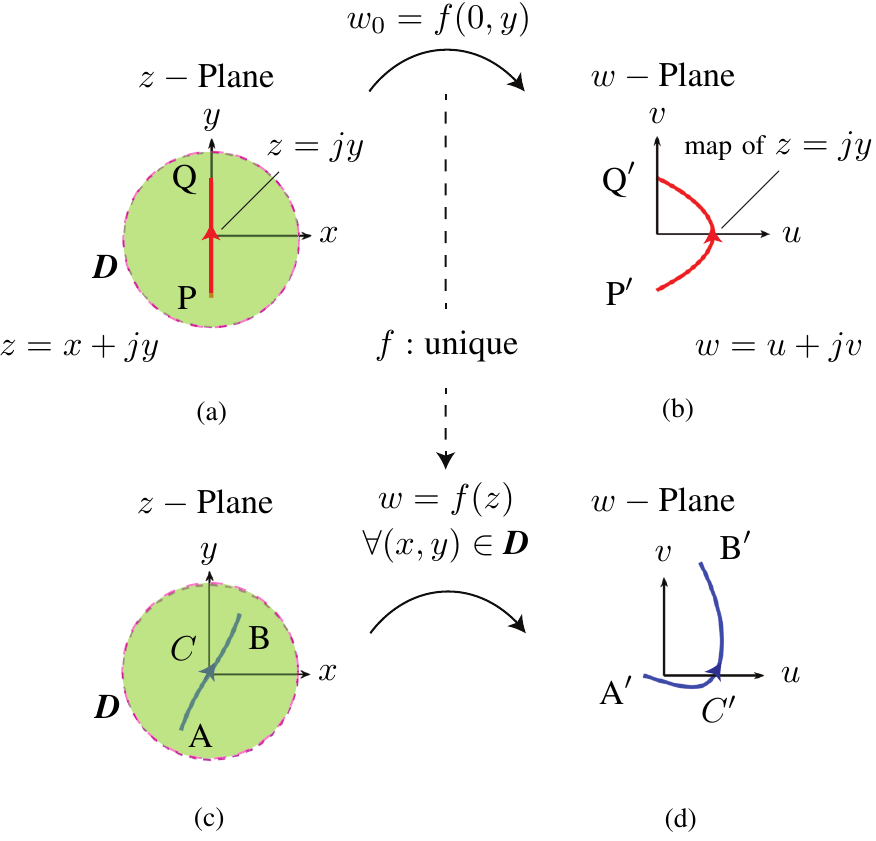}
	\caption{Illustration of the analytic continuation theorem. The green disk represents the domain of analycity of the function $f(\cdot)$. (a)~Line $z=jy$ (or $x=0$) in the $z$-plane within the analyticity domain \emph{\textbf{D}}. (b)~Map of this line into the $w$-plane. (c)~Arbitrary curve $C$ within the domain \emph{\textbf{D}} of the $z$-plane. (d)~Map of $C$ into the $w$-plane.}\label{fig:analytic_cont}
\end{figure}

Let us illustrate this theorem with the example of Fig.~\ref{fig:analytic_cont}, which depict a function $w=f(z)$ that maps the complex plane $z=x+jy$ into the complex plane $w=u+jv$. In this example, it is assumed that the mapping of the straight line $z=jy$ within \textbf{D} in Fig.~\ref{fig:analytic_cont}(a) is known as the curve plotted in Fig.~\ref{fig:analytic_cont}(b). If we can construct the function $f(\cdot)$, for instance expressed as a polynomial expansion, fitted to the data $[y,u,v]$, this function is \emph{unique} over the entire domain of analycity, and can therefore be used to map \emph{any} curve, $C$, in the $z$ plane, represented in Fig.~\ref{fig:analytic_cont}(c), into its image in the $w$ plane, as illustrated in Fig.~\ref{fig:analytic_cont}(d).

Let us further precise this explanation by considering a specific analytic mapping function $f(z)$ that is known in closed-form, as the mapping functions for the problems in Sec.~\ref{sec:spec_examples}. Consider the function $f(z)=1+z+z^2$, for which the image of the line $z=jy$ (or $x=0$) in Fig.~\ref{fig:analytic_cont}(a) is readily available as
\begin{equation}\label{eq:example}
w_0=f(0,y)=f(y)=(1-y^2)+jy,
\end{equation}
which corresponds in fact to the curve plotted in Fig.~\ref{fig:analytic_cont}(b) as $u=1-y^2$ and $v=y$ versus the parameter $y$. Since we know, from the analytic continuation theorem, that the function $f(z)$ between the $z$ and $w$ planes is unique, we can replace its argument $(0,y)$ or $y$ in~\eqref{eq:example} by any complex value $z=x+jy$ and still find the correct mapping. In particular, we may substitute $y=-jz$, which yields
\begin{align}\label{eq:example_f}
f(-jz)&=f(z)=f(x,y) \nonumber \\&=\left[1-(-jz)^2\right]+j(-jz) \nonumber \\&=1+z+z^2 \nonumber \\&=(x^2-y^2+x+1)+j(2xy+y) \nonumber \\&=u(x,y)+jv(x,y),
\end{align}
where the fourth equality indeed retrieves the original general analytic function $f(z)$ and the final result properly reduces to~\eqref{eq:example} for $z=(0,y)$. So, analytic continuity has allowed us to infer the general function $f(z)$, and hence the image of any curve $C$ of the $z$ plane in Fig.~\ref{fig:analytic_cont}(c), as illustrated in Fig.~\ref{fig:analytic_cont}(d)

Note that the mapping functions in Sec.~\ref{sec:spec_examples} have \emph{not} been found in this manner. Rather, these results were safely derived via the corresponding wave equations. However, we now know that this was an unnecessary complication. For instance, the general functions $\gamma=g^{-1}(\omega)$ of the fourth row in Tab.~\ref{tab:examples_analytic_sol} can alternatively be found by simply substituting $\omega_\text{r}=\omega_\text{r}+j\omega_\text{i}$ in the first row! But we shall next consider the more practical situations of mapping functions without analytical solution.

\subsection{General Mapping using Polynomial Expansion}\label{sec:proposed_method}

We shall apply now the procedure described in Sec.~\ref{sec:general_problem}, consolidated by the knowledge acquired in Sec.~\eqref{sec:act} that the mapping function is unique, due to its analycity established in Sec.~\ref{sec:anal_sol}. For the sake of simplicity, we shall restrict our attention to the determination of the function $\omega=g(\beta)$ (eigen-mode analysis computation), the reciprocal function $\gamma=g^{-1}(\omega_\text{r})$ (driven-mode analysis computation) being obtainable in an analoguous manner. Thus,

\begin{enumerate}
	\item First, we run an eigen-mode simulation of the problem at hand so as to compute the complex angular frequency $\omega(\beta)=\omega_\text{r}(\beta)+j\omega_\text{i}(\beta)$, and record the numerical result, i.e., $[\beta,\omega_\text{r},\omega_\text{i}]$.  
	\item Then, we build the function $g(\cdot)$ as the polynomial expansion 
	 \begin{align}\label{eq:omega_fun_beta}
	 \omega &=g(\beta) \nonumber \\
	 		&= \omega_\text{r}(\beta)+j\omega_\text{i}(\beta) \nonumber \\&= \sum_{m=0}^{M}A_m(\beta-\beta_0)^m+j\sum_{n=0}^{N}B_n(\beta-\beta_0)^n,
	 \end{align}
	 where ($M$,$A_m$) and ($N$,$B_n$) are the best fitted-polynomial degrees and coefficients to the functions $[\beta,\omega_\text{r}]$ and $[\beta,\omega_\text{i}]$, respectively\footnote{The MATLAB function $\textsf{polyfit(x,y,n)}$ returns the coefficients of the polynomial of degree $\textsf{n}$ that best fits the numerical function $[\textsf{x},\textsf{y}]$.} around the analytical point $\beta_0$.
	\item Next, we replace the argument in the function $g(\beta)$ by $-j\gamma$ in the dispersion relation~\eqref{eq:omega_fun_beta}, which consequently transforms into
	\begin{align}\label{eq:omega_fun_gamma}
	\omega&=g(\gamma) \nonumber \\ &= \sum_{m=0}^{M}A_m(-j\gamma-\beta_0)^m+j\sum_{n=0}^{N}B_n(-j\gamma-\beta_0)^n.
	\end{align}
	which relates $\omega$ and $\gamma$.
	\item Finally, we compute the Max($M$,$N$) complex $\gamma=\alpha+j\beta$ roots of the equation 
	\begin{equation}\label{poly_gamma}
	\sum_{m=0}^{M}A_m(-j\gamma-\beta_0)^m+j\sum_{n=0}^{N}B_n(-j\gamma-\beta_0)^n-\omega_\text{r}=0,
	\end{equation}
	which provides the sought after map $\gamma(\omega_\text{r})=\alpha(\omega_\text{r})+j\beta(\omega_\text{r})$\footnote{We naturally search for complex roots $\gamma$ whose imaginary parts are close to $\beta_0$. For example, the function $\textsf{cxroot(FUN,z0)}$ in MATLAB finds a complex root of the function $\textsf{FUN}$ close to an initial guess $\textsf{z0}$.}.   
\end{enumerate}

Note that an accurate estimation of the polynomial $\omega(\beta)$ in~\eqref{eq:omega_fun_beta} requires a sufficient number and range of data points of complex $\omega$ versus $\beta$ values. For instance, finding the correct curve between $\text{P}'$ and $\text{Q}'$ in the $w$-plane in Fig.~\ref{fig:analytic_cont}(b) (similarly in the $\omega$-plane) requires a sufficiently large number of points between the two points P and Q and a sufficiently extended range [P,~Q] in the $z$-plane in Fig.~\ref{fig:analytic_cont}(a) (similarly in the $\omega$-plane). Practically, a simple convergence analysis indicates when the number and range of points is sufficient.

Moreover, the point $\beta_0$ must be chosen in a region where the function $\omega(0+j\beta)$ is analytic, i.e., where the function $\omega$ and all its derivatives are continuous. Let us consider for example the case of the unbounded lossy medium, where $\omega(\gamma)$ is given by Eq.~\eqref{eq:complex_frequency_gamma_UM}. In this case, $\gamma=j\sigma/2\epsilon \nu$ is the branch point and we may choose the branch cut $\beta<\sigma/2\epsilon \nu$, which requires $\beta_0>\sigma/2\epsilon \nu$. In the case of a medium with relatively low loss, $\beta_0$ may be safely set to approximately zero.   

Finally, the proposed technique is not restricted to 1-D (periodic) structures; it can be readily generalized to 2-D and 3-D (periodic) structures. For example, for a 2-D periodic structure with periods $p_1$ and $p_2$ along the direct lattice vectors ${\textbf{a}}_1$ and ${\textbf{a}}_2$, we only need to run the eigen-mode analysis along the three segments of the irreducible Brillouin zone, formed by the reciprocal vectors ${\textbf{b}}_1$ and ${\textbf{b}}_2$, i.e., along the spectral paths $\Gamma-X$, $X-M$ and $M-\Gamma$ for a square lattice or $\Gamma-K$, $K-R$ and $R-\Gamma$ for a hexagonal lattice~\cite{Joannopoulos1995photonic}. Once the complex frequencies, $\omega_p$, have been computed along the 3 $k_p$ paths ($p=1,2,3$), we can use the same algorithm as for the 1-D case to find $\omega_p=g(\gamma_p)$, with $\omega$, $\beta$ and $\gamma$ respectively replaced by $\omega_p$, $k_p$ and $\gamma_p$.

\section{Validation and Illustrations}\label{sec:val_ill}
We shall now illustrate the general mapping technique established in Sec.~\ref{sec:General_solution} for four practical examples and validate the results by either closed-form analytical solutions or driven-mode analysis full-wave results.   
\subsection{Lossy Medium}
First, we consider the lossy medium, analytically treated in Sec.~\ref{sec:lossy_med}. From the complex frequency $\omega(\beta)$ data given by Eq.~\eqref{eq:complex_frequency}, we will see if the proposed mapping technique is consistent with the analytical solution, given by~\eqref{eq:propagation_constant}. 

\begin{figure}[h!]
	\begin{center}
		\subfigure[] {
			\includegraphics[width=0.45\columnwidth]{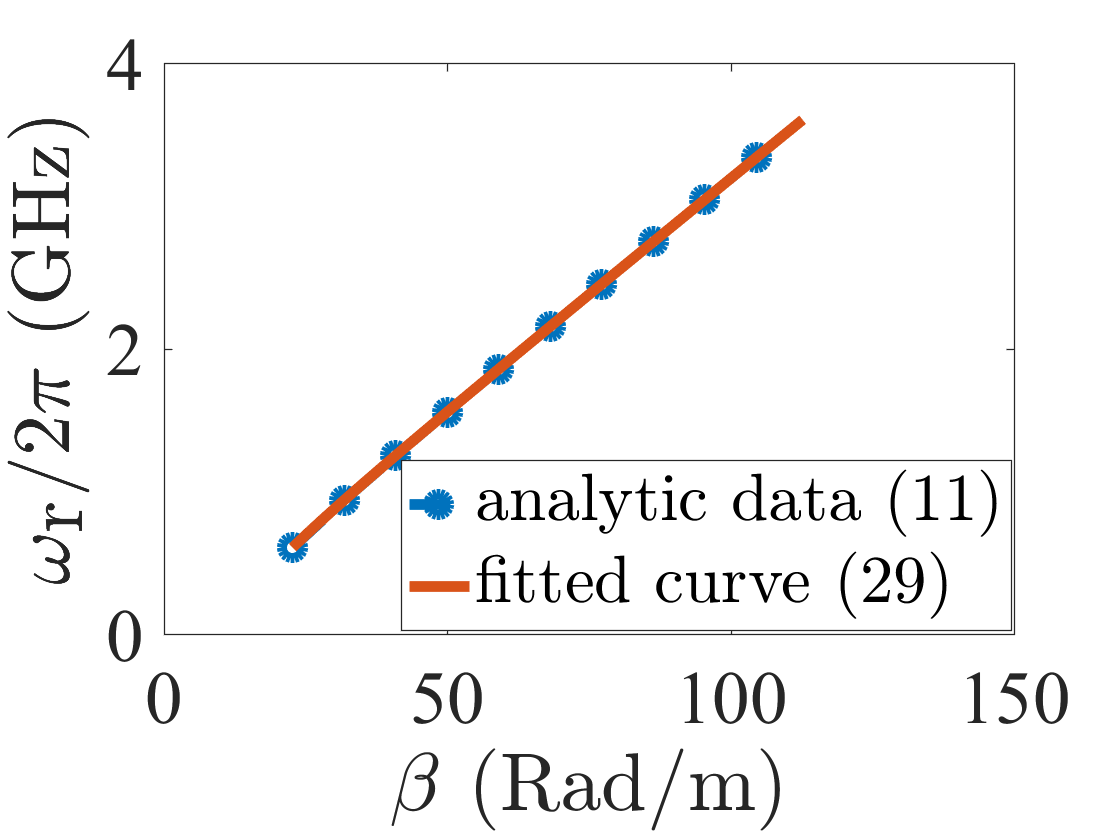}
		}
		\subfigure[] {
			\includegraphics[width=0.45\columnwidth]{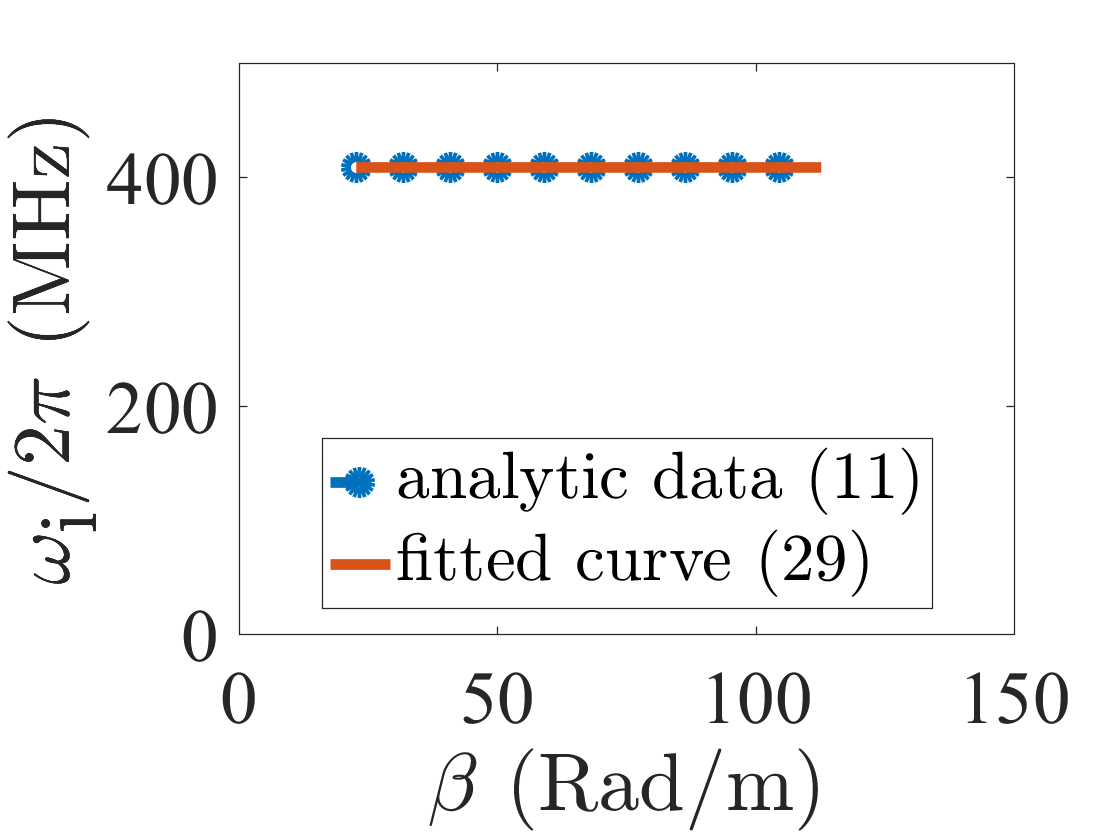}
		}
		\subfigure[] {
			\includegraphics[width=0.45\columnwidth]{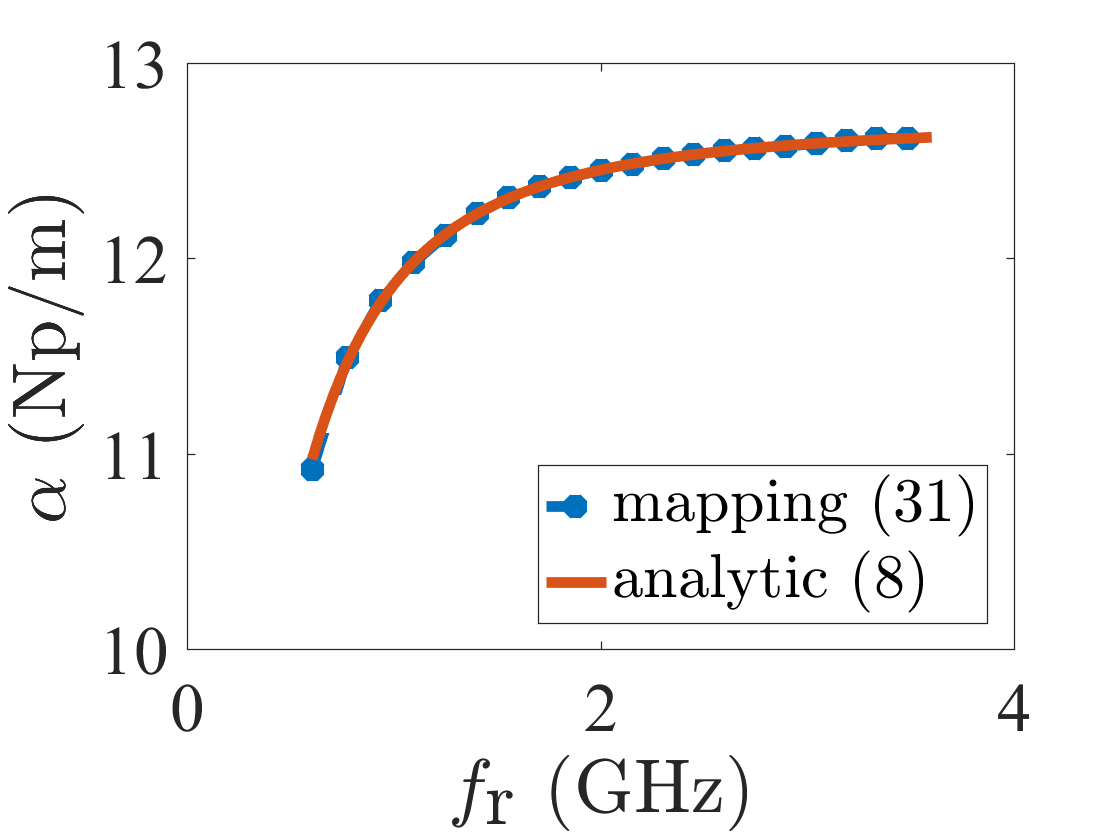}
		}
		\subfigure[] {
			\includegraphics[width=0.45\columnwidth]{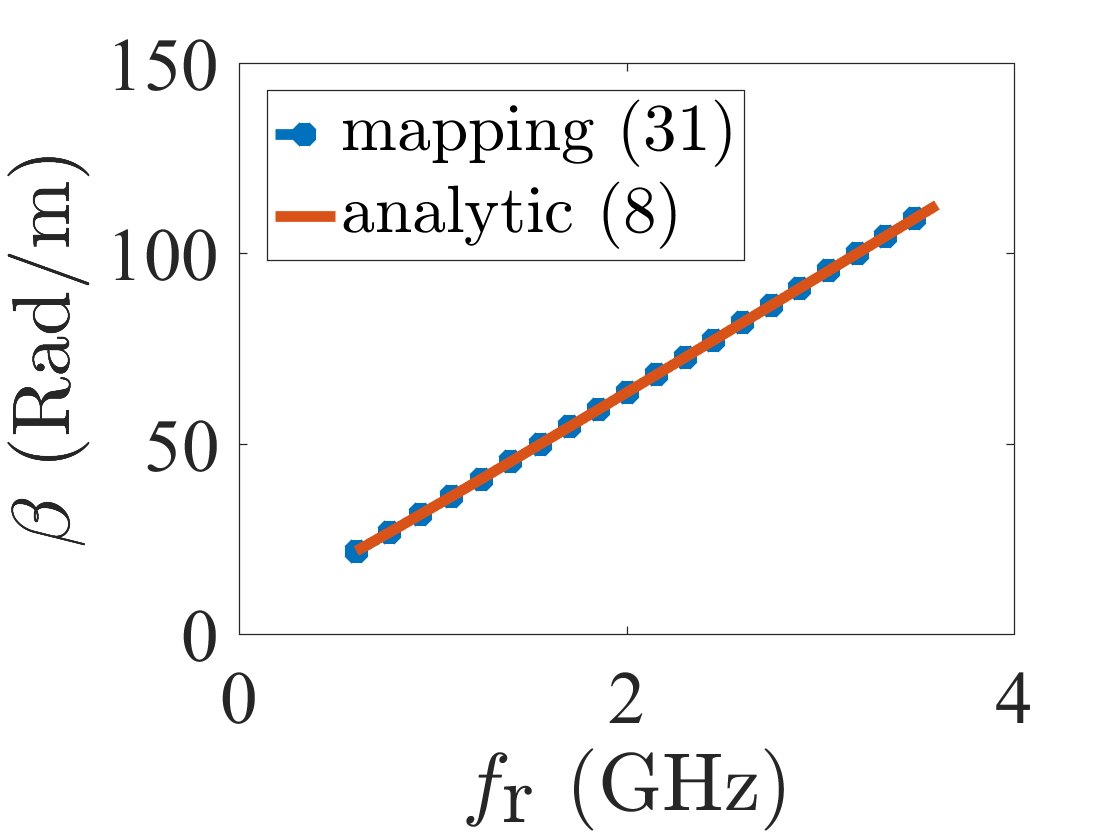}
		}
		\caption{General mapping for an unbounded lossy dielectric medium with relative permittivity $\epsilon_\text{r}=2.2$ and conductivity $\sigma=0.1~\textrm{S/m}$.
		(a)~$\omega_\text{r}(\beta)$. (b)~$\omega_\text{i}(\beta)$. (c)~$\alpha(f_\text{r})$. (d)~$\beta(f_\text{r})$. The fitting polynomials of $\omega_\text{r}(\beta)$ and $\omega_\text{i}(\beta)$ have degrees 10 and 0, respectively, and $\beta_0$ is set $22.7 > \sigma/2\epsilon \nu \approx 12.7$ rad/m.}
		\label{Fig:mapping_UBM}
	\end{center}
\end{figure}

Figures~\ref{Fig:mapping_UBM}(a) and (b) show the $\omega_\text{r}$ and $\omega_\text{i}$ data using Eq.~\eqref{eq:complex_frequency} and their fitted curves/polynomials using Eq.~\eqref{eq:omega_fun_beta} as functions of $\beta$, respectively. Figures~\ref{Fig:mapping_UBM}(c) and (d) respectively show the $\alpha$ and $\beta$ estimated by the mapping technique~Eq.\eqref{poly_gamma} and the analytical formulation~\eqref{eq:propagation_constant}, and a close agreement between the two is observed.  

\subsection{Dielectric-filled Metallic Waveguide}
Second, we consider the rectangular waveguide in Sec.\ref{sec:met_wg}, filled with a lossy dielectric. From the complex frequency $\omega(\beta)$ data given by Eq.~\eqref{eq:complex_frequency_WG}, we will see if the proposed mapping technique is in agreement with the analytical solution, given by Eq.~\eqref{eq:propagation_constant_WG}. 

\begin{figure}[h!]
	\begin{center}
		\subfigure[] {
			\includegraphics[width=0.45\columnwidth]{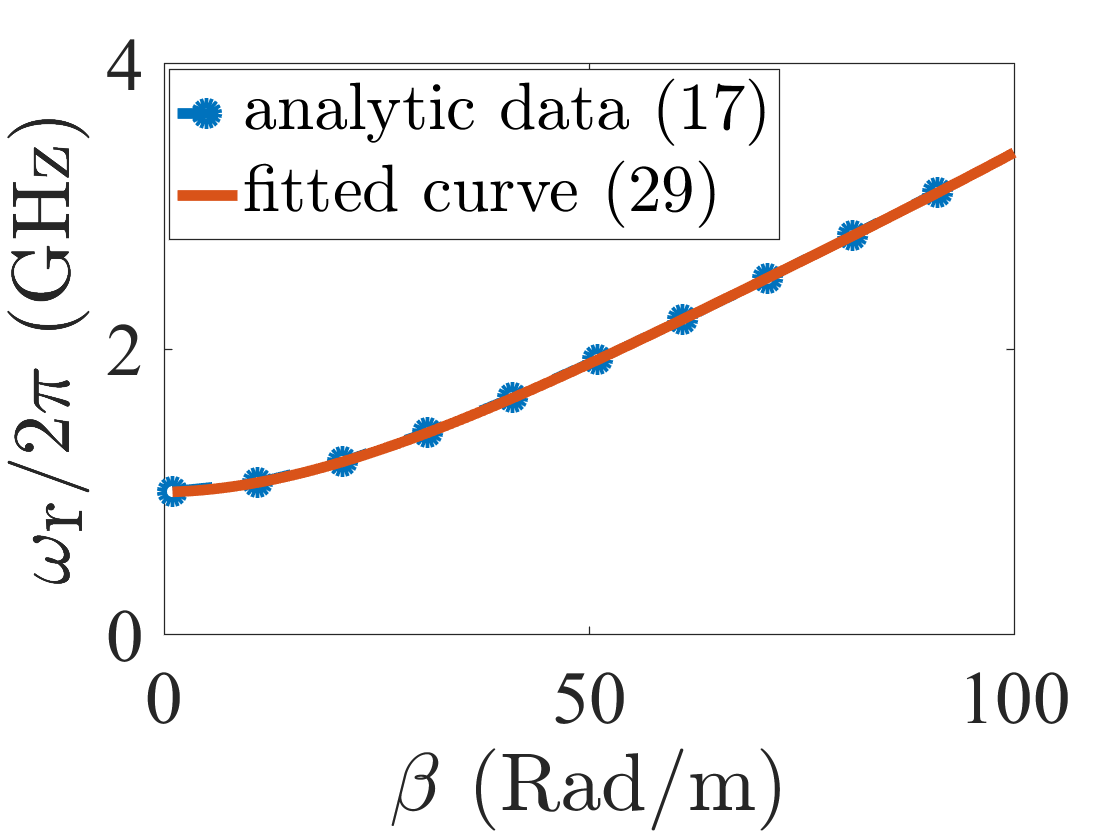}
		}
		\subfigure[] {
			\includegraphics[width=0.45\columnwidth]{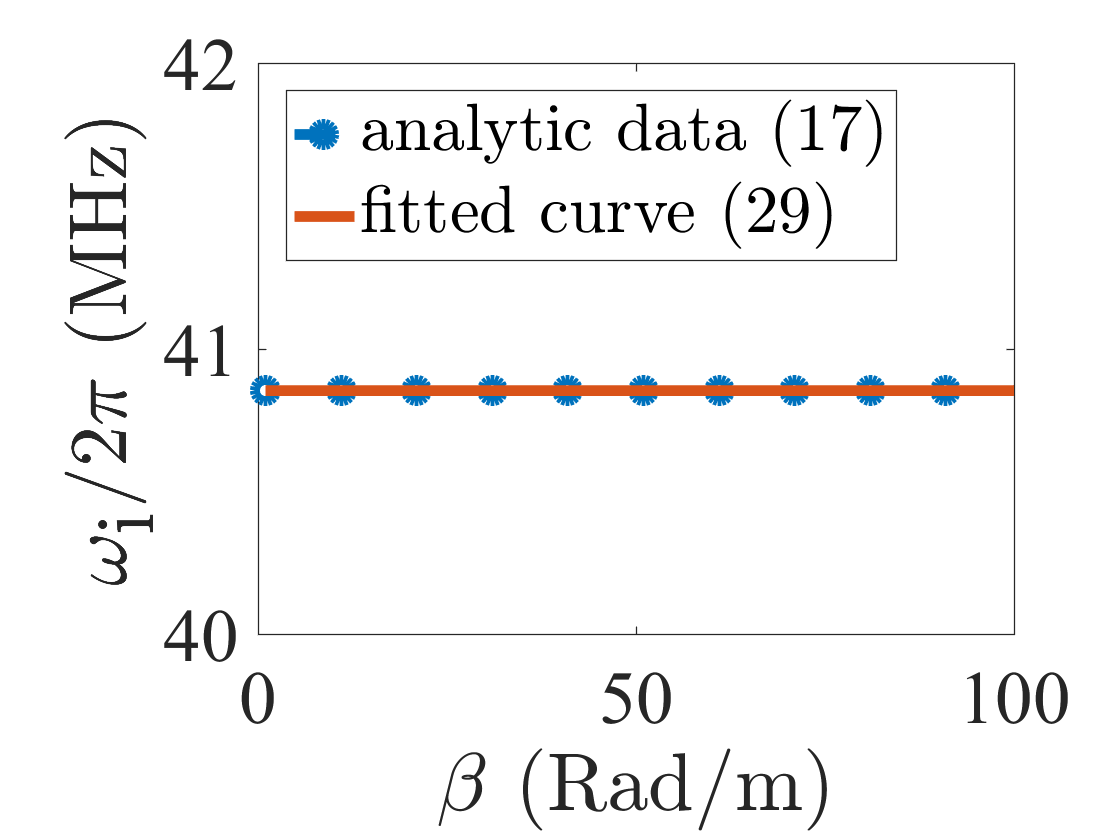}
		}
		\subfigure[] {
			\includegraphics[width=0.45\columnwidth]{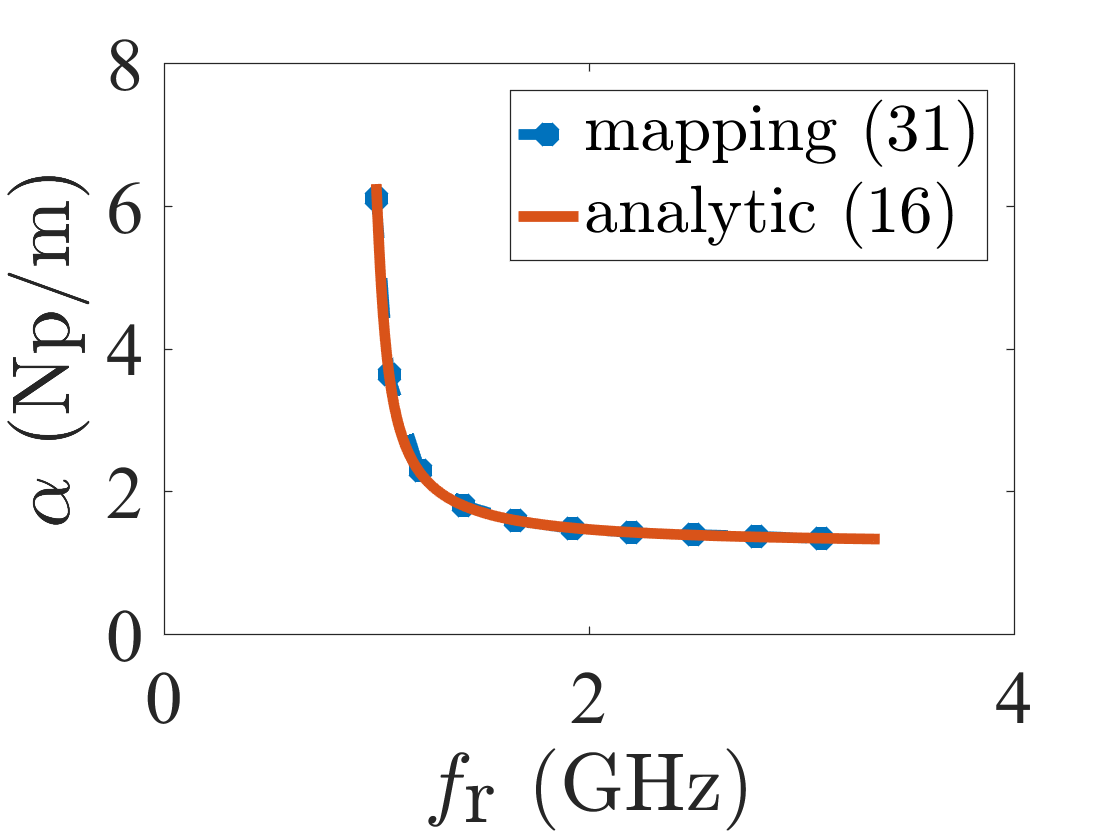}
		}
		\subfigure[] {
			\includegraphics[width=0.45\columnwidth]{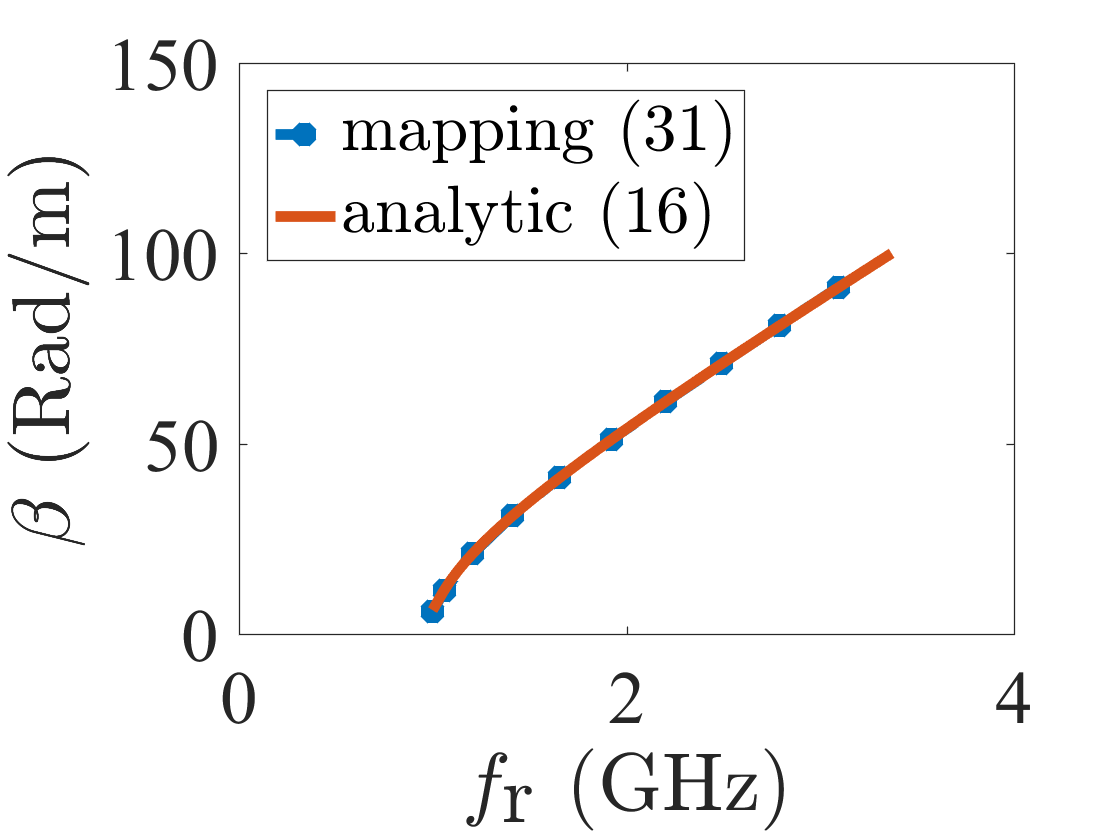}
		}
		\caption{General mapping for a metallic rectangular waveguide filled with a lossy dielectric with $\epsilon_\text{r}=2.2$ and $\sigma=0.01~\textrm{S/m}$ and width $a=\lambda_0/2\sqrt{\epsilon_\text{r}}$ where $\lambda_0=30~\text{cm}$. (a)~$\omega_\text{r}(\beta)$. (b)~$\omega_\text{i}(\beta)$. (c)~$\alpha(f_\text{r})$. (d)~$\beta(f_\text{r})$. The fitting polynomials of $\omega_\text{r}(\beta)$ and $\omega_\text{i}(\beta)$ have degrees 4 and 0, respectively and \mbox{$\beta_0=0$}.}
		\label{Fig:mapping_WG}
	\end{center}
\end{figure}

Figures~\ref{Fig:mapping_WG}(a) and (b) respectively show the $\omega_\text{r}$ and $\omega_\text{i}$ data using Eq.~\eqref{eq:complex_frequency_WG} and their fitted polynomials as functions of $\beta$. Figures~\ref{Fig:mapping_UBM}(c) and (d) respectively show the $\alpha$ and $\beta$ estimated by the mapping technique and the analytical formulation~\eqref{eq:propagation_constant_WG} where perfect agreement between the two is observed.  

\subsection{1-D Photonic Crystal}
\begin{figure}[h!]
	\includegraphics[width=0.9\columnwidth]{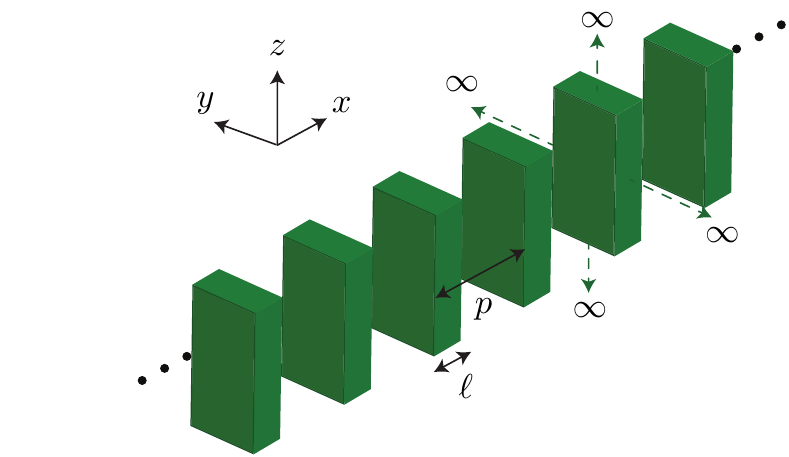}
	\caption{Photonic crystal consisting of lossy dielectric slabs with $\epsilon_\text{r}=4$ and $\sigma=0.01~\textrm{S/m}$, the periodicity $L=3~\text{cm}$ and thickness $\ell=L/2$.}\label{fig:periodic_slab}
\end{figure}

Figure \ref{fig:periodic_slab} shows a 1-D periodic photonic crystal~\cite{Joannopoulos1995photonic} whose dispersion relation is given by (Appendix~\ref{sec:derivation_dd_ps})
\begin{equation}\label{eq:dis_rel_per_slab}
\text{det}
\begin{pmatrix}
\text{e}^{-jk\ell}   &~~\text{e}^{jk\ell}    &-\text{e}^{-jk_0\ell}      & -\text{e}^{jk_0\ell}      \\
\text{e}^{-jk\ell}   &-\text{e}^{jk\ell}     &-\zeta\text{e}^{-jk_0\ell} &~~\zeta\text{e}^{jk_0\ell} \\
\text{e}^{-\gamma L} &~~\text{e}^{-\gamma L} &-\text{e}^{-jk_0L}         & -\text{e}^{jk_0L}         \\
\text{e}^{-\gamma L} &-\text{e}^{-\gamma L}  &-\zeta\text{e}^{-jk_0L}    &~~\zeta\text{e}^{jk_0L}
\end{pmatrix}=0,
\end{equation}
where $k=k_0 \sqrt{\epsilon_\text{rc}}$, with $k_0=\omega/c$ ($c$: speed of light in vacuum), $\epsilon_\text{rc}={\epsilon_\text{r}-j\sigma/\omega\epsilon_0}$ and $\zeta=1/\sqrt{\epsilon_\text{rc}}$. 

\begin{figure}[h!]
	\begin{center}
		\subfigure[] {
			\includegraphics[width=0.45\columnwidth]{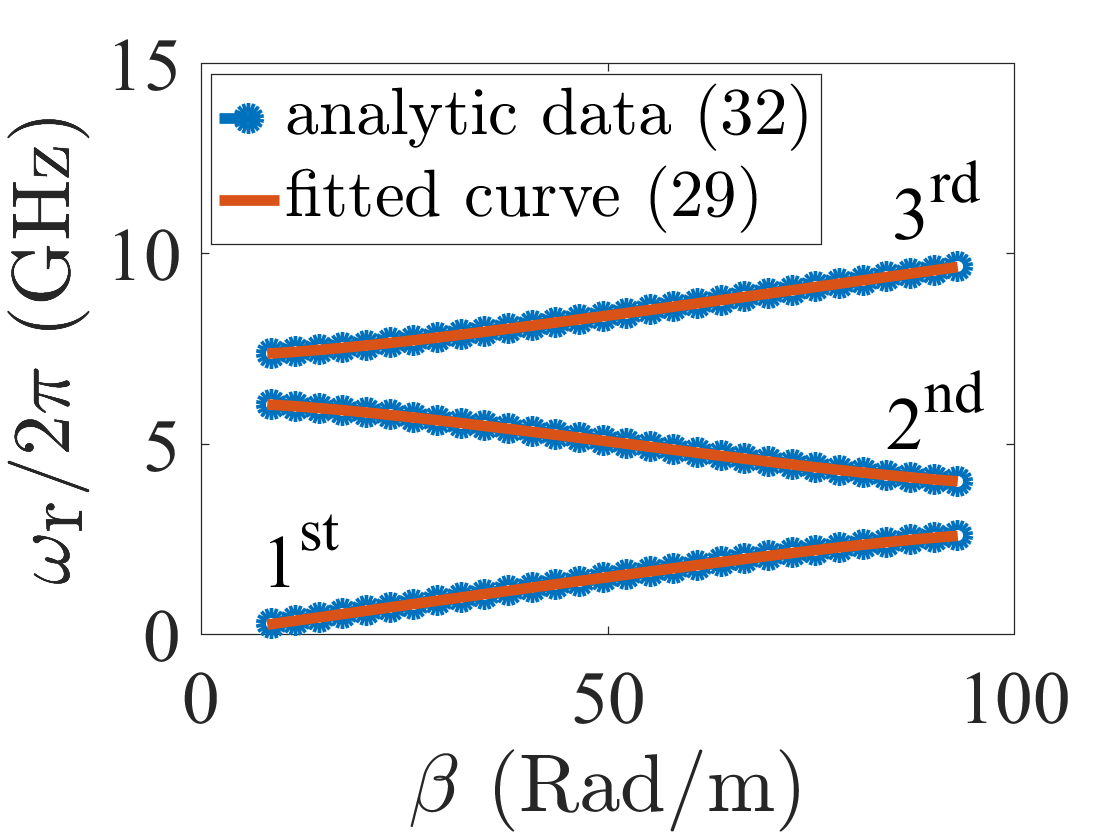}
		}
		\subfigure[] {
			\includegraphics[width=0.45\columnwidth]{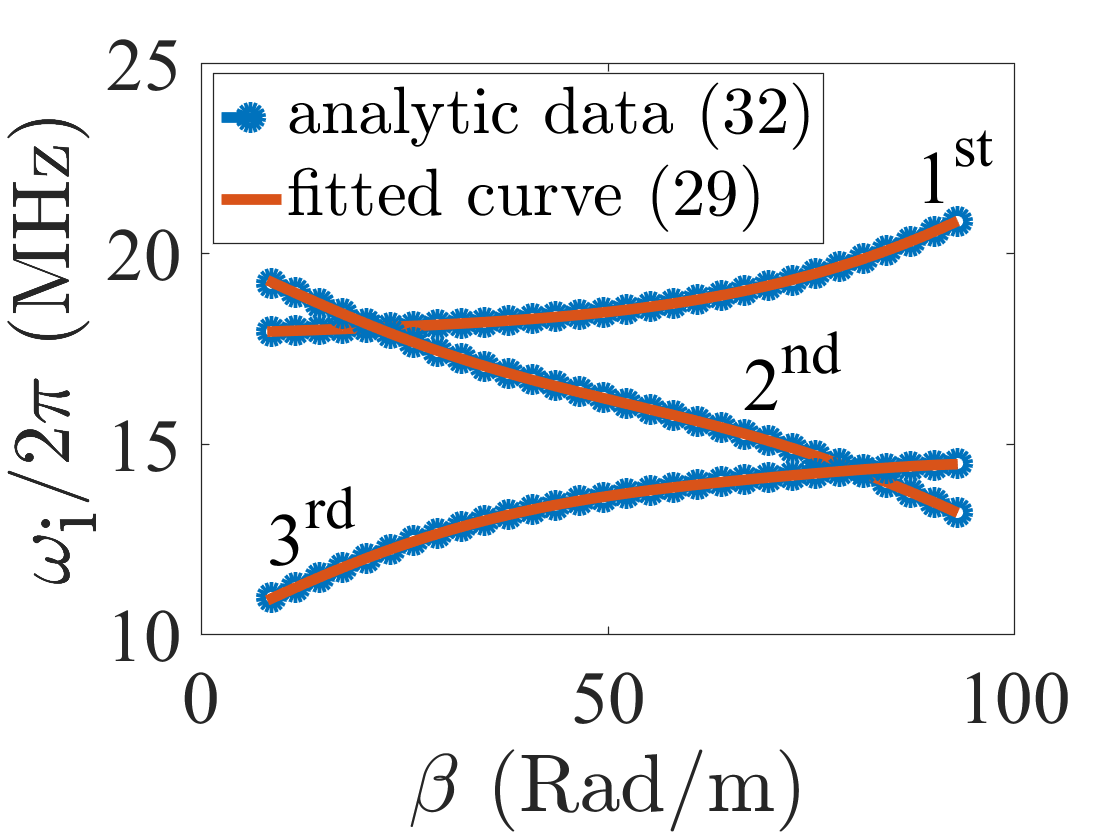}
		}
		\subfigure[] {
			\includegraphics[width=0.45\columnwidth]{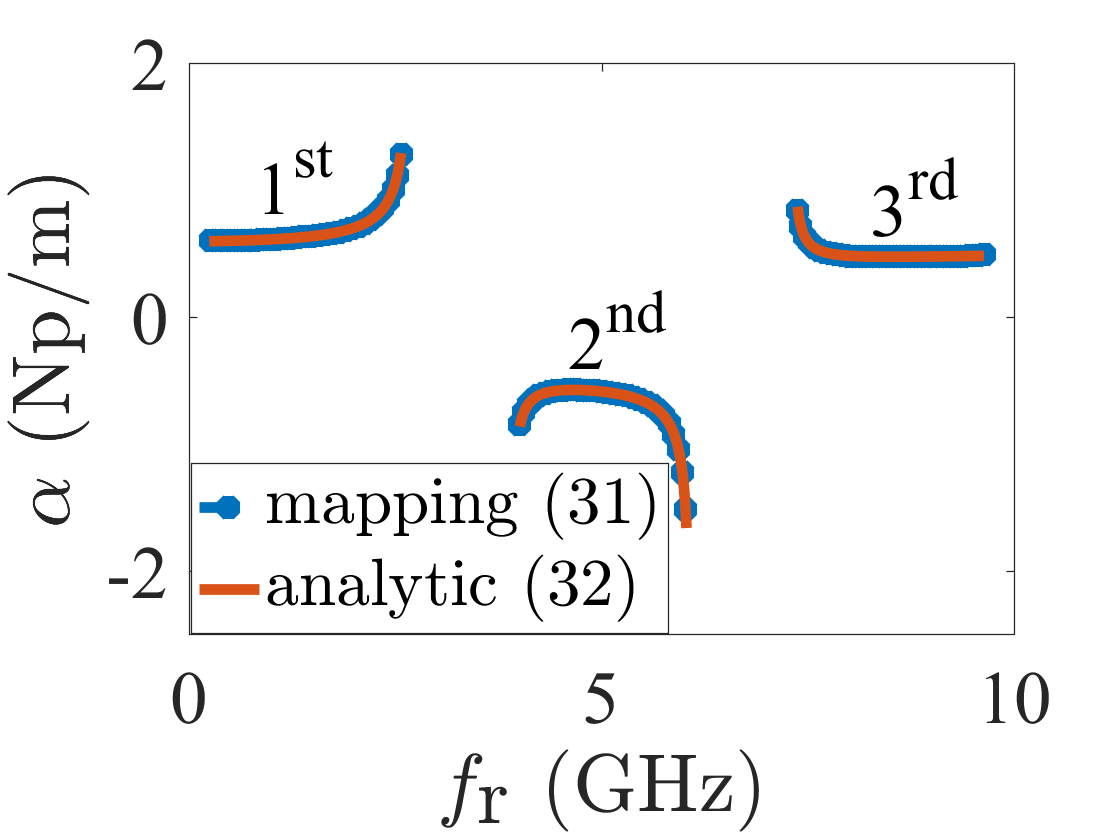}
		}
		\subfigure[] {
			\includegraphics[width=0.45\columnwidth]{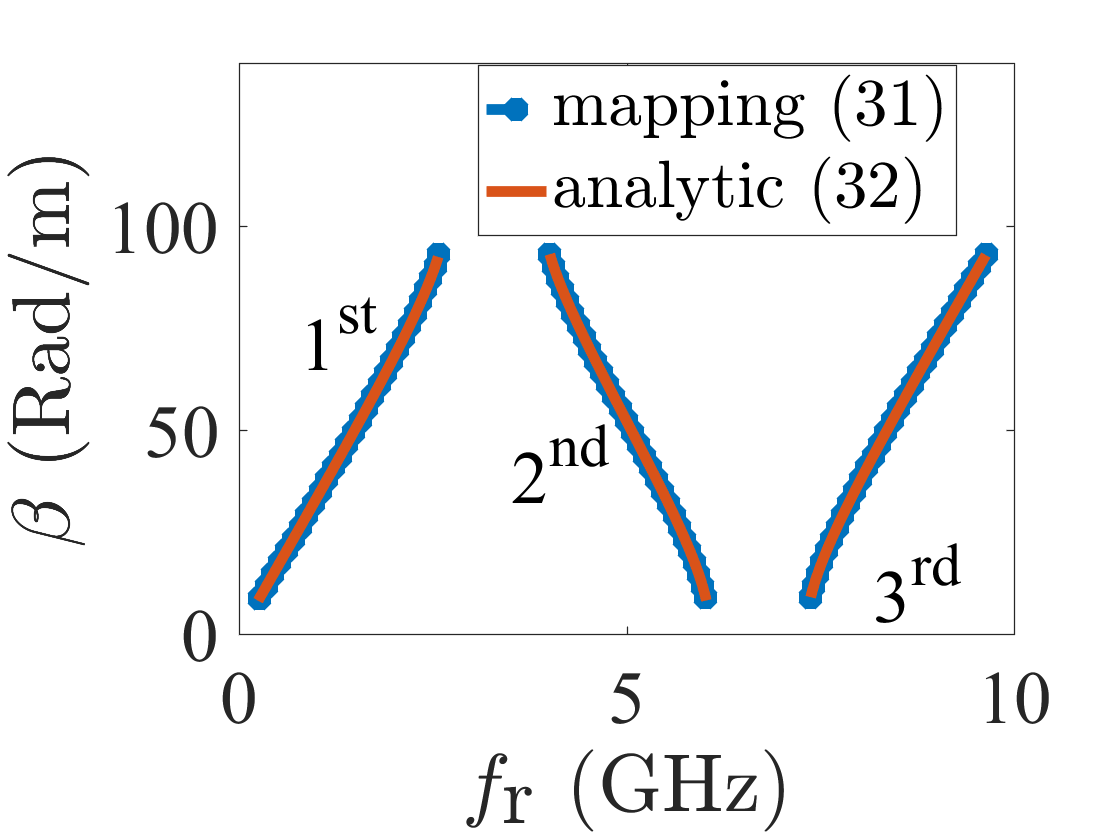}
		}
		\caption{General mapping for the 1-D photonic crystal in Fig.~\ref{fig:periodic_slab} solutions for the first 3 space harmonics. (a)~$\omega_\text{r}(\beta)$. (b)~$\omega_\text{i}(\beta)$. (c)~$\alpha(f_\text{r})$. (d)~$\beta(f_\text{r})$. The polynomials of $\omega_\text{r}(\beta)$ and $\omega_\text{i}(\beta)$ both have the degree 5, and $\beta_0=0$.}
		\label{Fig:mapping_PS}
	\end{center}
\end{figure}

Equation~\eqref{eq:dis_rel_per_slab} provides an analytical (but not closed-form) mapping function from the $\omega$-plane into the $\gamma$-plane and vise versa. Setting $\gamma=j\beta$ in Eq.~\eqref{eq:dis_rel_per_slab} and finding the complex roots $\omega$ for which the determinant is zero yields the real and imaginary frequencies as functions of $\beta$. Figures~\ref{Fig:mapping_PS}(a) and (b) respectively show the $\omega_\text{r}$ and $\omega_\text{i}$ data and their fitted polynomials as functions of $\beta$ for the first 3 space harmonics. 

Next, we apply the proposed mapping technique and compare the results with the analytical ones. The analytical results are again calculated by Eq.~\eqref{eq:dis_rel_per_slab} in which we set the $\omega$ to be purely real $\omega_\text{r}$ and look for the complex roots $\gamma$ for which the determinant is zero. Figures~\ref{Fig:mapping_PS}(c) and (d) respectively show the $\alpha$ and $\beta$ estimated by the mapping technique and the analytical formulation based on~\eqref{eq:dis_rel_per_slab} where again perfect agreements between the two are observed.
\subsection{Leaky-Wave Antenna}
\begin{figure}[h!]
	\subfigure[] {
		\includegraphics[width=\columnwidth]{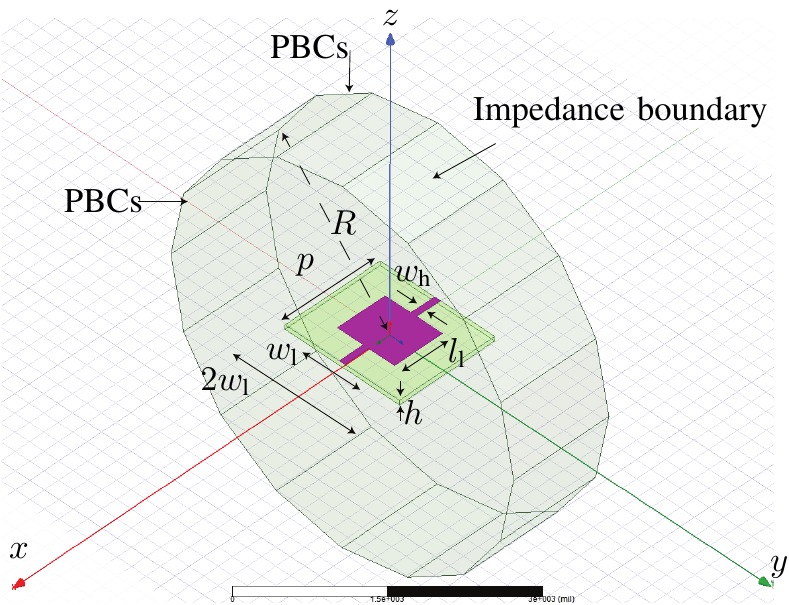}
}
\subfigure[] {
	\centering
	\includegraphics[width=\columnwidth]{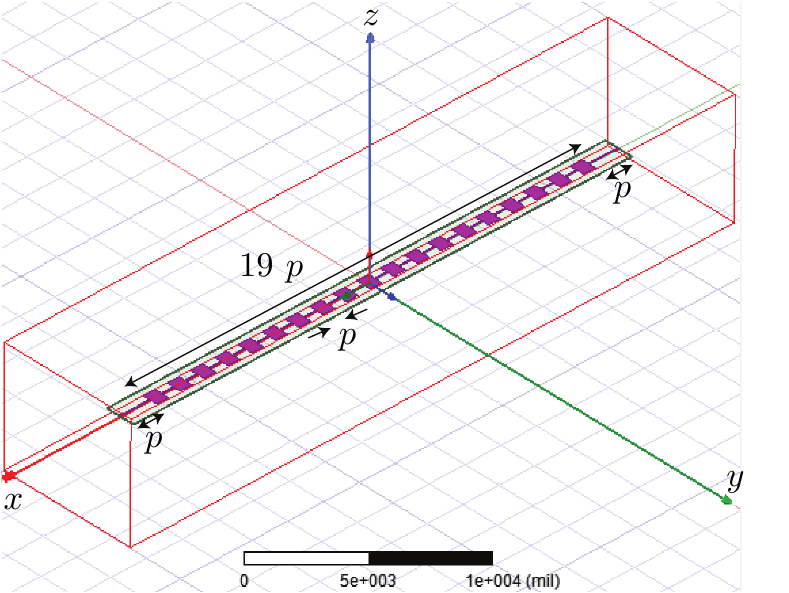}
}
\caption{Periodic LWA problem. The antenna center frequency is set to $f_\textrm{0}=5.8~\textrm{GHZ}$ and the unit cell dimensions are $p=33.64~\text{mm}$, $l_\text{l}=p/2$, $w_\text{l}=20~\text{mm}$ and $w_\text{h}=2~\text{mm}$~\cite{otto2011transmission}. The (lossless) substrate has a relative permittivity of $\epsilon_\text{r}=2.2$ and a height of $h=1.5~\text{mm}$. Ansys HFSS (a)~eigen-mode and (b)~driven-mode analysis setups. In the eigen-mode analysis the unit cell is placed within 16-segment polygonal cylinder with the radius $R=6~\text{cm}$ and assigned PBCs to the front and back faces and assigned surface impedance of $120\pi$ to the peripheral faces.}\label{fig:LWA}
\end{figure}

Let us now consider the problem of a series-fed patch (SFP) periodic leaky-wave antenna (LWA), shown in Fig.~\ref{fig:LWA}, for which no analytical solution exist and full-wave numerical analysis is necessary~\cite{otto2011transmission}. Figures~\ref{fig:LWA}(a) and~\ref{fig:LWA}(b) show the unit cell setup of the eigen-mode analysis and the entire LWA 19-cell structure with the microstrip transmission line ports in the driven-mode analysis, respectively.
As shown in Fig.~\ref{fig:LWA}(a), we place the unit cell inside a polygonal cylinder with assigned PBCs to the front and back faces and assigned surface impedance of $120\pi$ to the peripheral faces and a large enough\footnote{The radius $R \gg \lambda_0$ is large enough so that the field at the cylinder boundary can be approximated by a plane wave. Here, $\lambda_0\approx 5.2~\text{cm}$ at $f_\textrm{c}=5.8~\textrm{GHz}$.} radius to simulate radiation into free-space. 

\begin{figure}[h!]
	\begin{center}
		\subfigure[] {
			\includegraphics[width=0.45\columnwidth]{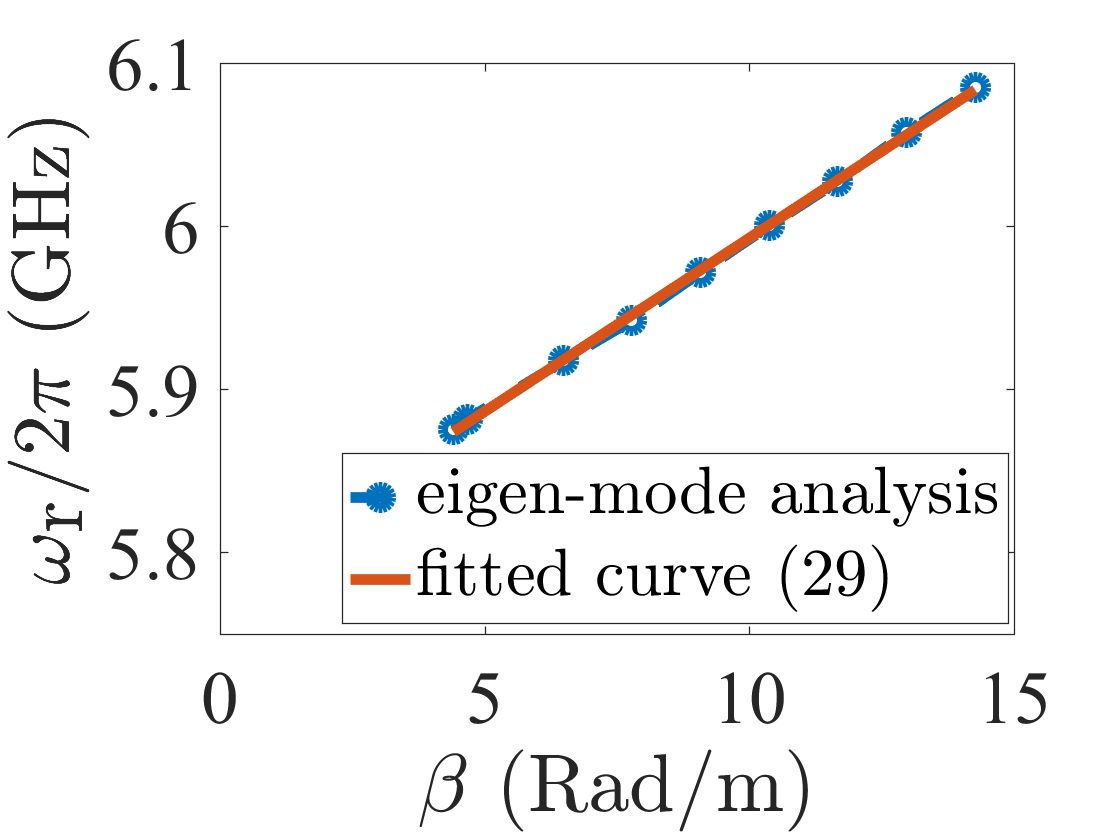}
		}
		\subfigure[] {
			\includegraphics[width=0.45\columnwidth]{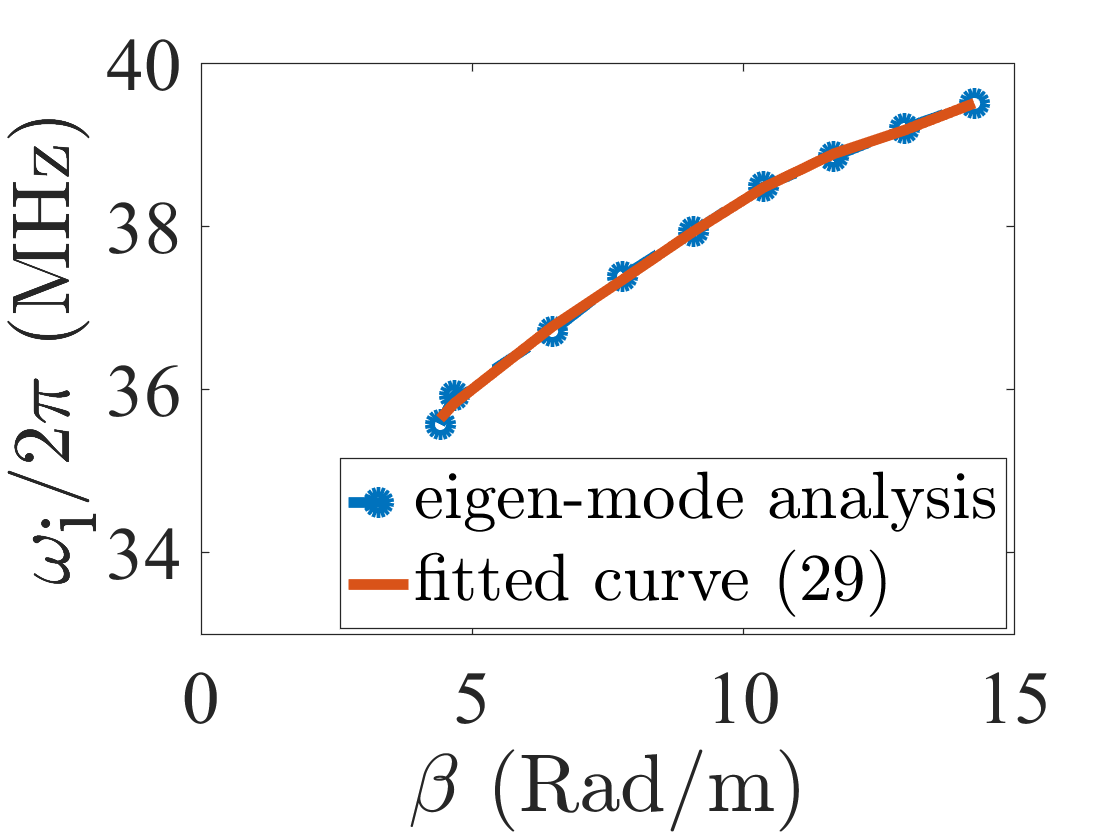}
		}
		\subfigure[] {
			\includegraphics[width=0.45\columnwidth]{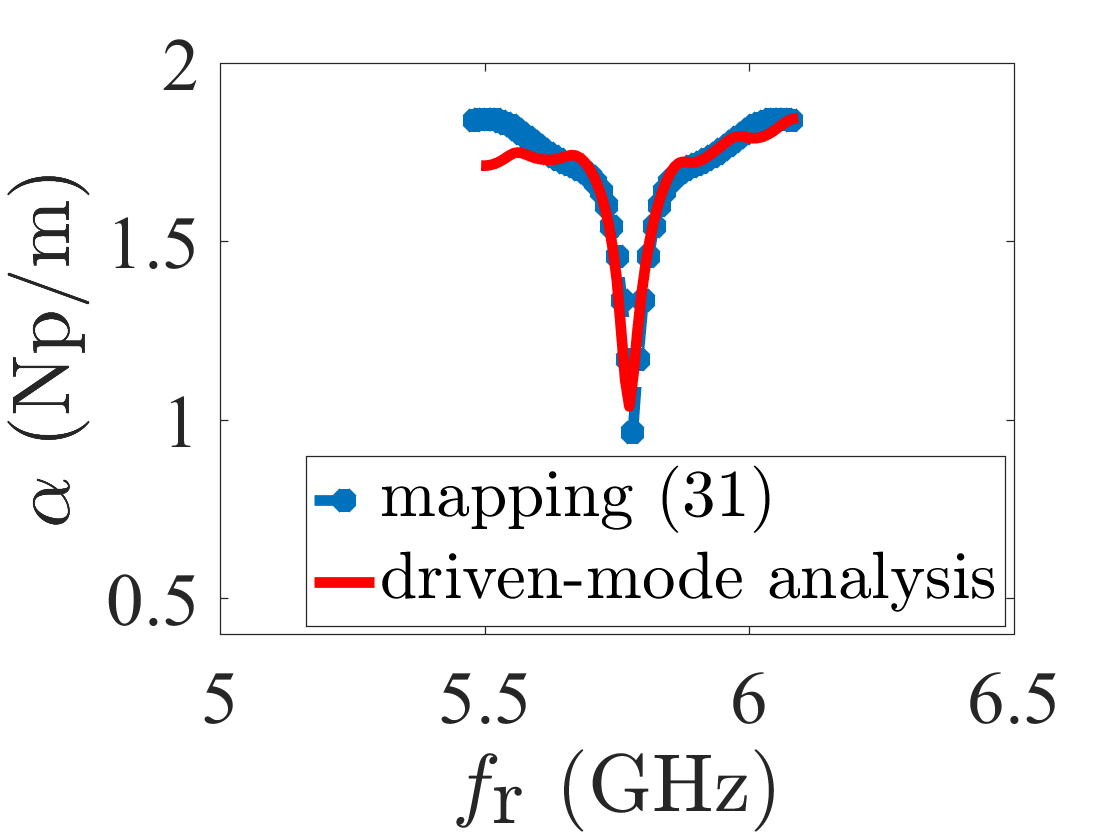}
		}
		\subfigure[] {
			\includegraphics[width=0.45\columnwidth]{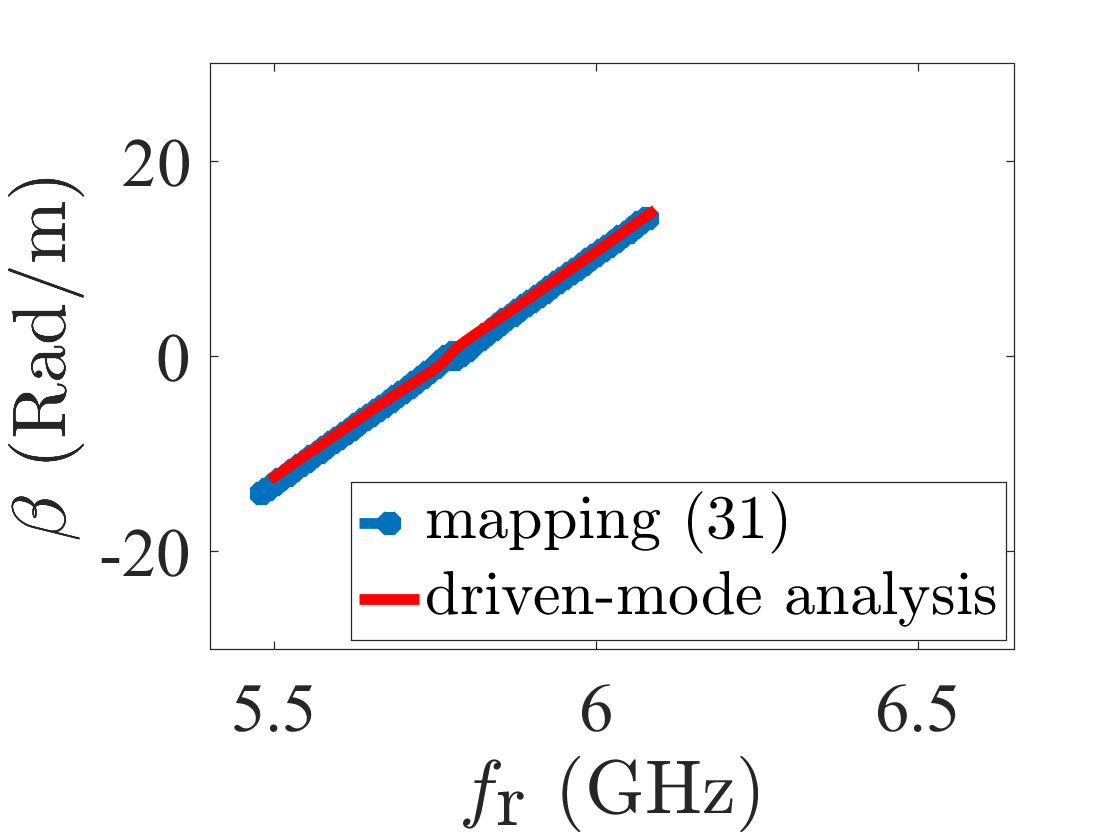}
		}
		\caption{General mapping for the leaky-wave antenna in Fig.~\ref{fig:LWA}. (a) $\omega_\text{r}(\beta)$. (b) $\omega_\text{i}(\beta)$. (c) $\alpha(f_\text{r})$. (d) $\beta(f_\text{r})$. The polynomials of $\omega_\text{r}(\beta)$ and $\omega_\text{i}(\beta)$ have degrees 1 and 5, respectively, and $\beta_0=0$.}
		\label{Fig:mapping_LWA}
	\end{center}
\end{figure}

Figures~\ref{Fig:mapping_LWA}(a) and (b) show the real and imaginary frequencies, $\omega_\text{r}$ and $\omega_\text{i}$ given by the eigen-mode analysis\footnote{The Ansys HFSS eigen-mode solver provides $\omega$ in terms of phase difference $\phi$ between the front and back faces of the polygonal cylinder shown in Fig.~\ref{fig:LWA}(a). The phase constant is then given by $\beta=-\phi/p$.} and their corresponding fitted polynomials as functions of $\beta$. 
We then apply the proposed mapping technique and compare the results with solutions of the driven-mode analysis of the Ansys HFSS\footnote{ Ansys HFSS gives the scattering (S)-parameters of the two-terminal LWA. After deembedding the S-parameters of the periodic patches from the response of the entire structure, patches and the microstrip transmission lines, we calculate the $ABCD$ matrix parameters of the periodic patches and then calculate $\gamma$, given by $\gamma=\cosh^{-1}A$ \cite{pozar2011microwave}.}. Figures~\ref{Fig:mapping_LWA}(c) and (d) respectively show the $\alpha$ and $\beta$ estimated by the mapping technique and the Ansys HFSS driven-mode analysis where great agreement between the two is observed.
\section{Conclusion}\label{sec:Dis_Con}
We introduced a new general method for mapping the complex spatial and temporal frequencies of an arbitrary electromagnetic structure, with and without analytical solution. The method is based on the analyticity of the physical mapping function and uses a fitted polynomial expansion to build this function, which is guaranteed by analytic continuity to be unique. This is a fundamental method, which is expected to find many applications in Electromagnetics.


\appendices
\numberwithin{figure}{section}
\numberwithin{equation}{section}

\section{Derivation of Equation~\eqref{eq:dis_rel_per_slab}}\label{sec:derivation_dd_ps}
Let us assume a $z$-polarized plane wave propagating along the $x$ direction. The electromagnetic fields in the slab, ($E_z$, $H_y$), and in free-space, ($E_{z0}$, $H_{y0}$), regions of the periodic unit cell, are given in terms of forward and backward propagating waves, with ($A$,$B$) and ($C$,$D$) being unknown coefficients, as
\begin{subequations}\label{eq:fields}
\begin{equation}
\begin{pmatrix} E_z \\ H_y \end{pmatrix} = A \begin{pmatrix} 1 \\ {-1}/{\eta} \end{pmatrix} \text{e}^{-jkx} + B \begin{pmatrix} ~~1 \\ {1}/{\eta} \end{pmatrix} \text{e}^{jkx},
\end{equation}
\begin{equation}
\begin{pmatrix} E_{z0} \\ H_{y0} \end{pmatrix} = C \begin{pmatrix} 1 \\ {-1}/{\eta_0} \end{pmatrix} \text{e}^{-jk_0x} + D \begin{pmatrix} ~~1 \\ {1}/{\eta_0} \end{pmatrix} \text{e}^{jk_0x},
\end{equation}
\end{subequations}
where $\eta=\eta_0/\sqrt{\epsilon_\text{rc}}$, $\epsilon_\text{rc}={\epsilon_\text{r}-j\sigma/\omega\epsilon_0}$, $\eta_0=\sqrt{\mu_0/\epsilon_0}$, $k=k_0 \sqrt{\epsilon_\text{rc}}$, $k_0=\omega/c$ and $c$ is the speed of light in vacuum.

Applying the continuity boundary conditions $(E_z=E_{z0})|_{x=\ell}$ and $(H_y=H_{y0})|_{x=\ell}$, 
\begin{subequations}\label{eq:BCs}
	\begin{align}
	& A \begin{pmatrix} 1 \\ -1/\eta \end{pmatrix} \text{e}^{-jk\ell} + B \begin{pmatrix} ~~1 \\ 1/\eta \end{pmatrix} \text{e}^{jk\ell}=\nonumber \\
	& C \begin{pmatrix} 1 \\ -1/\eta_0 \end{pmatrix} \text{e}^{-jk_0\ell} + D \begin{pmatrix} ~~1 \\ 1/\eta_0 \end{pmatrix} \text{e}^{jk_0\ell}
	\end{align}
	and next the periodic boundary conditions $\text{e}^{-\gamma L} E_z|_{x=0}=E_{z0}|_{x=L}$ and $\text{e}^{-\gamma L} H_y|_{x=0}=H_{y0}|_{x=L}$, given by
	\begin{align}
	& \text{e}^{-\gamma L} \left[ A \begin{pmatrix} 1 \\ -1/\eta \end{pmatrix} + B \begin{pmatrix} ~~1 \\ 1/\eta \end{pmatrix} \right] =\nonumber \\
	& C \begin{pmatrix} 1 \\ -1/\eta_0 \end{pmatrix} \text{e}^{-jk_0 L} + D \begin{pmatrix} ~~1 \\ 1/\eta_0 \end{pmatrix} \text{e}^{jk_0 L},
	\end{align}
\end{subequations}
results in the matrix equation
\begin{equation}\label{eq:coefficient_slab}
\begin{pmatrix}
\text{e}^{-jk\ell}   &~~\text{e}^{jk\ell}    &-\text{e}^{-jk_0\ell}      & -\text{e}^{jk_0\ell}      \\
\text{e}^{-jk\ell}   &-\text{e}^{jk\ell}     &-\zeta\text{e}^{-jk_0\ell} &~~\zeta\text{e}^{jk_0\ell} \\
\text{e}^{-\gamma L} &~~\text{e}^{-\gamma L} &-\text{e}^{-jk_0L}         & -\text{e}^{jk_0L}         \\
\text{e}^{-\gamma L} &-\text{e}^{-\gamma L}  &-\zeta\text{e}^{-jk_0L}    &~~\zeta\text{e}^{jk_0L}
\end{pmatrix} 
\begin{pmatrix}
A    \\
B    \\
C    \\
D 
\end{pmatrix}
=
\begin{pmatrix}
0    \\
0    \\
0    \\
0 
\end{pmatrix},
\end{equation}     
where $\zeta=\eta/\eta_0=1/\sqrt{\epsilon_\text{rc}}$. For a non-trivial solution, the coefficient matrix must have a zero determinant, which results into Eq.~\eqref{eq:dis_rel_per_slab}.  
\bibliography{References}{}
\bibliographystyle{IEEEtran}
\end{document}